\documentclass[a4paper,11pt]{article}
\usepackage{jheppub} 
\usepackage{lineno}
\usepackage{tabularx}
\usepackage{array}
\usepackage{comment}

\newcommand{\be}{\begin{equation}}
\newcommand{\ee}{\end{equation}}
\newcommand{\bea}{\begin{eqnarray}}
\newcommand{\eea}{\end{eqnarray}}

\newcommand{\nn}{\nonumber}

\newcommand{\ket}[1]{\left| #1 \right>}
\newcommand{\bra}[1]{\left< #1 \right|}

\newcommand{\ba}{\begin{eqnarray}}
\newcommand{\ea}{\end{eqnarray}}

\newcommand{\ep}{\epsilon}
\newcommand{\fq}{\mathfrak{q}}

 \def\ep{\epsilon}
 \def\commutator#1#2{[#1, #2]}

\usepackage{xcolor}
\definecolor{amethyst}{rgb}{0.54, 0.17, 0.89}
\definecolor{coral}{rgb}{1.0, 0.3, 0.4}
\definecolor{green}{rgb}{0.05,0.5,0.1}

\newcommand{\beq}{\begin{equation}}
\newcommand{\eeq}{\end{equation}}
\newcommand{\beqn}{\begin{eqnarray}}
\newcommand{\eeqn}{\end{eqnarray}}
\newcommand{\cN}{\mathcal{N}}

\title{Wigner negativity in Krylov space and emergent semiclassicality}

\author[a,b,c,h]{Vijay Balasubramanian,}
\author[d,e,f]{Pawel Caputa,}
\author[g]{Onkar Parrikar,}
\author[g]{Vivek Singh}

\affiliation[a]{David Rittenhouse Laboratory, University of Pennsylvania,
209 S. 33rd Street, Philadelphia, PA 19104, USA}
\affiliation[b]{Santa Fe Institute,
1399 Hyde Park Road, Santa Fe, NM 87501, USA}
\affiliation[c]{Theoretische Natuurkunde, Vrije Universiteit Brussel,
Pleinlaan 2, B-1050 Brussels, Belgium}
\affiliation[d]{
The Oskar Klein Centre and Department of Physics, Stockholm University, AlbaNova, 106 91 Stockholm, Sweden}
\affiliation[e]{Yukawa Institute for Theoretical Physics, Kyoto University, Kitashirakawa Oiwakecho, Sakyo-ku, Kyoto 606-8502, Japan}
\affiliation[f]{Faculty of Physics, University of Warsaw, Pasteura 5, 02-093 Warsaw, Poland}
\affiliation[g]{Department of Theoretical Physics, 
Tata Institute of Fundamental Research, 1 Homi Bhabha Road, Mumbai 400005, India}
\affiliation[h]{International Centre for Theoretical Sciences-TIFR,
Shivakote, Hesaraghatta Hobli, Bengaluru North 560089, India}

\preprint{YITP-26-82, TIFR/TH/26-19}

\abstract{We propose that the Krylov basis gives a semiclassical representation of dynamics in general large-$N$, complex, many-body systems. As a probe of this semiclassicality, we study the growth of Wigner negativity -- a measure of the complexity of classical simulation -- under time evolution in Krylov space in  several solvable models. We begin with 2d CFTs, initially in either the vacuum  or the thermofield double state on a line excited by a primary operator. In both cases,   Wigner negativity remains an $O(1)$ constant and does not grow at late times, indicating approximately classical dynamics in the Krylov basis. We then study random matrix theory with the  maximally entangled state between two copies as the initial state. For general one-cut matrix models, we argue that  Wigner negativity in the Krylov basis grows as $t^{1/2}$ at large $O(1)$ times but does not scale with the Hilbert space dimension, thus indicating semiclassical dynamics in Krylov space. Finally, in the double-scaled SYK model, we find an approximately classical phase (constant negativity) at early times and a semiclassical phase ($t^{1/2}$ growth) at late times. In all these examples,  Wigner negativity either remains  constant or grows slowly, demonstrating emergent semiclassicality of dynamics in Krylov space.} 

\begin{document}

\maketitle

\parskip=10pt
\section{Introduction}
Quantum mechanics is crucially different from classical mechanics because quantum states can be superposed. As a result, they are vectors in a Hilbert space whose dimension is exponentially large in the system size. Consider, for example, that to describe the time evolution of a single classical particle one must keep track of just two numbers, namely its position and momentum, while  quantum mechanically the state is specified by the particle's wavefunction with respect to some basis. If the state of the system remains ``simple'' in the computational basis under time evolution -- where by simple we mean in terms of an $O(1)$ number of wavefunction coefficients -- then the time evolution is tractable on a classical computer. However, if the state becomes ``complex'' with respect to the computational basis under time evolution, in the sense that its wavefunction spreads out over an exponentially large number of basis states, then we must  track  exponentially many wavefunction coefficients, which is intractable for a classical computer in the limit that number of degrees of freedom $N$ diverges. In this case, it would be helpful to find, if possible, a new basis where the initial state and its time evolution remain simple in the sense of remaining effectively confined to an $O(1)$ subspace at any $O(1)$ time in the $N\to \infty$ limit. If such a basis can be found, then we say that the dynamics in that basis provides a {\it low-complexity effective description} of the quantum system valid at $O(1)$ times.

The Krylov basis  \cite{viswanath1994recursion} is an ideal candidate for such an effective description  (see  \cite{Nandy:2024htc,Rabinovici:2025otw,Baiguera:2025dkc} for  reviews). Given an initial state $\ket{\psi_0}$ and Hamiltonian $H$, the Krylov basis is obtained by orthonormalizing the ordered set of states $\left\{\ket{\psi_0}, H\ket{\psi_0}, H^2\ket{\psi_0},\cdots \right\}$ via the Gram-Schmidt process. Since the initial state $\ket{\psi_0}$ is one of the basis vectors, it has a very simple wavefunction in this basis, namely $(1,0,0,\cdots,0)$. Furthermore, the Hamiltonian in the Krylov basis takes a tri-diagonal form. The non-zero components along and just off the diagonal comprise the {\it Lanczos spectrum} of the theory, and lead to  effectively local propagation of the wavefunction along the Krylov chain. It  was shown in \cite{Balasubramanian:2022tpr} that the spread of the wavefunction across the Krylov basis, quantified by {\it spread complexity}, is the slowest possible in any basis, at least at early times. Further, in \cite{2024JHEP...05..264B} this argument was generalized to show that the Krylov basis (with a suitable choice of phases) also minimizes  \emph{Wigner negativity} of the spreading state at early times.  

Wigner negativity is computed from the Wigner function, a basis-dependent quasi-probability representation of quantum states on an appropriately defined  phase space.  The Wigner function  cannot always be understood as a classical probability distribution because it can take negative values, and the amount of negativity can intuitively be thought of as a measure of the complexity of classically simulating a quantum circuit preparation of the state. More formally,  Wigner negativity is a stabilizer/magic monotone, which means that it is an operationally meaningful measure of  non-stabilizerness with respect to a given choice of basis \cite{Veitch_2014}.  In this sense, Wigner negativity measures the {\it classicality} of a state as described in a given basis. The perturbative arguments in \cite{Balasubramanian:2022tpr, 2024JHEP...05..264B} thus indicate that the Krylov basis simultaneously provides the {\it most compact} (and hence least complex) and the {\it most classical} description of quantum time evolution at the same time.\footnote{Krylov space techniques, are partly for this reason, routinely used in numerical time evolution for tensor network states (see \cite{Paeckel:2019yjf} for a review).}  If the dynamics is largely confined for some period of time to a part of the  Krylov basis, we could  hope to build a low-complexity, semiclassical  effective description of the system by restricting attention to the Hilbert subspace spanned by that subset. 

Evidence along these lines comes from Random Matrix Theory (RMT). Consider a chaotic Hamiltonian modeled by a random matrix of dimension $D=e^S$ drawn from the Gaussian unitary ensemble (GUE) and with a generic initial state. Time evolution in this system with respect to a generic choice of basis develops an exponentially large $O(e^{S/2})$ amount of Wigner negativity within an $O(1)$ amount of time in units set by the width of the spectral density \cite{Basu:2025mmm}. It thus becomes intractable to classical simulation as $S\to\infty$. In sharp contrast,  the negativity in the Krylov basis grows at most as a power law -- i.e., $O(\sqrt{t})$ -- at large times $t \gg 1$ \cite{Basu:2025mmm}. 
So for all $O(1)$ times, by which we mean timescales that do not grow parametrically with the system size $S$, the negativity remains of $O(1)$.  In this regime dynamics in the Krylov basis remains semiclassical. Additionally, the wavefunction starts out localized on a single Krylov basis element and then spreads essentially ballistically across the basis \cite{Balasubramanian:2022tpr,Balasubramanian:2022dnj}.  This implies that at $O(1)$ times, the state will be well-described   by $O(1)$ non-zero coefficients in the Krylov basis. Taken together, these observations show that, over any  duration that does not scale with the system size, the Krylov basis provides a low-complexity, semiclassical effective description -- it minimizes the size of the subspace required to approximate the wavefunction well, and at the same time it  permits  tractable classical simulation. This low-complexity \emph{Krylov effective description} of the dynamics generated by a random Hamiltonian is valid at any fixed $O(1)$ time in the $S\to \infty$ limit. However, at exponentially large times $t \sim e^S$, the negativity becomes exponentially large and saturates to a close-to-maximal value, implying that even the Krylov basis description of the state must be highly quantum  \cite{Basu:2025mmm}.

Here, we will extend the above ideas by studying the time evolution of Wigner negativity in Krylov space for various systems such as 2d Conformal Field Theories (CFTs), RMT, and the Double-Scaled SYK (DSSYK) model. In all these cases we will study dynamics in a large system size limit, and will treat Wigner negativity as a measure of semiclassicality with respect to a given choice of basis. Specifically we will be interested in the degree to which the time evolution of a classical initial state remains semiclassical with respect to this measure. In a large-but-finite dimensional system, a typical state will have negativity proportional to $\sqrt{D}$ with respect to a generic choice of basis, where $D$ is the dimension of the Hilbert space  \cite{Basu:2025mmm}. States with negativity scaling in this way are irreducibly quantum mechanical. By contrast, we say that a state is \emph{classical} with respect to a choice of basis if it has vanishing Wigner negativity in that basis. We will say that a state is \emph{semiclassical}  with respect to a given choice of basis if the Wigner negativity does not diverge in the $D\to \infty$ limit.
Likewise, Hamiltonian time evolution in this language is classical if it does not increase Wigner negativity in the given choice of basis. We will say that the dynamics is semiclassical if the negativity potentially grows but remains $O(1)$ at times of $O(1)$ and does not scale with the system size.  As $D \to \infty$,  a system starting in a classical state and subject to such a semiclassical evolution requires an infinite time to reach negativity scaling with system size. The purpose of this paper is to study these notions of semiclassicality with respect to a specific choice of basis, namely the Krylov basis. Indeed, as discussed above, the Krylov basis minimizes negativity growth at early times; in this paper we study negativity growth at finite and large but O(1) times. In Sec.~\ref{sec:prelim}, we begin by reviewing quantum dynamics in the Krylov basis, the definition of spread complexity, and the definition of Wigner negativity along with its operational significance.  We also define a  Krylov-Wigner function associated to evolution of a quantum state in the Krylov basis for infinite dimensional Hilbert spaces, and  relate the notions of semiclassicality discussed above to conventional notions in terms of saddle points in path integrals.

In Sec.~\ref{sec:2dCFT}, we analyze 2d CFTs with the initial state  taken to be either the vacuum  or the thermofield double (TFD) on a line, excited by a primary operator of dimension $\Delta$. In this case, the Hilbert space, and hence the Krylov basis, are infinite dimensional and the tridiagonalized Hamiltonian has asymptotically linearly growing Lanczos coefficients. As a result, under time evolution the wavefunction of such states spreads indefinitely on the Krylov chain. The associated spread complexity \cite{Balasubramanian:2022tpr}, which quantifies the dispersal of the wavefunction on the Krylov chain, grows quadratically in time in the vacuum case, and exponentially in the TFD case. By contrast,  in both cases Wigner negativity saturates to an $O(1)$ constant and does not grow with time when $\Delta > 1/2$.\footnote{For $\Delta \leq 1/2$,  negativity grows with time, but there is still a sense in which it grows much slower than spreading of the wavefunction would naively allow.} Thus, the complexity of classical simulation remains small even though the spread complexity in the Krylov basis grows rapidly (even exponentially in the thermal case), indicating an approximately \emph{classical} phase of time evolution.  This also implies that the early growth in complexity of the state is an essentially classical phenomenon in this case, and not a quantum one. 

In Sec.~\ref{sec:RMT}, we study the growth of Wigner negativity for  maximally entangled initial states of two copies in RMT. For the Gaussian unitary ensemble in the large system limit, the elements of the Lanczos spectrum turn out to be an $O(1)$ constant (independent of the Krylov depth), and as a result the spread complexity grows linearly with time. Likewise, we show that the Wigner negativity in the Krylov basis admits a finite limit as the matrix dimension increases, and grows dynamically as $\sqrt{t}$ at late times. (This improves upon the argument in \cite{Basu:2025mmm} that the growth is {\it at most}  $O(\sqrt{t})$.)  Thus, for times of $O(1)$ the negativity also remains of $O(1)$ indicating,  in the nomenclature we discussed above, a \emph{semiclassical} phase of time evolution that is also  within the purview of low-complexity effective theory.  In fact, we expect that this $O(\sqrt{t})$ growth is universal for any unitarily invariant ensemble where the large-$D$ density of energy eigenvalues is compactly supported on an interval. This stems from the fact that for any such random matrix ensemble, the Lanczos spectrum at sufficiently large but $O(1)$ Krylov depth resembles that of the GUE \cite{Kar:2021nbm, math_paper}. Finally, in Sec.~\ref{sec:DSSYK} we study Wigner negativity in the DSSYK model. In this case, we find that the dynamics is effectively classical (constant negativity) at early times (when the system explores the regime of Krylov depth $n \lambda \ll 1$\footnote{Here $\lambda$ is related to the q-deformation parameter via $\mathfrak{q}=e^{-\lambda}.$}), but at later times (when the system explores the regime $n\lambda \gg 1$) the negativity grows as $t^{1/2}$.

The authors of \cite{2024JHEP...05..264B} suggested that in the context of AdS/CFT, we should think of gravity as  a low-complexity effective description of the boundary quantum dynamics in the sense described above.\footnote{There are also other notions of quantum complexity, such as circuit complexity, which have been widely studied in the gravity context, see \cite{Stanford:2014jda, Jefferson_2017, Chapman_2018, Susskind:2018pmk, Brown_2018, Balasubramanian:2019wgd, Balasubramanian:2021mxo} and references therein.}  This idea is borne out in the DSSYK model \cite{Berkooz:2018jqr,Berkooz:2018qkz, Lin:2022rbf, Rabinovici:2023yex} where the dual q-deformed semiclassical JT gravity theory emerges from the boundary perspective in the Krylov basis. In particular,  the length of wormholes in the emergent gravity turns out to be equal to the spread complexity of the DSSYK thermofield double state \cite{Lin:2022rbf, Rabinovici:2023yex}. In the gravity description the length of these wormholes grows forever, but at exponentially late times the spread complexity saturates \cite{Balasubramanian:2022tpr,Balasubramanian:2024lqk, Balasubramanian:2025xkj}.  A possible reason for this discrepancy is that the state is becoming highly quantum mechanical so that a semiclassical notion of length simply does not apply. This is indeed what happens from the perspective of Wigner negativity, which saturates in RMT \cite{2024JHEP...05..264B, Basu:2025mmm} to an exponentially large value at times of $O(D)$, indicating that the effective description in the Krylov basis is irreducibly quantum.

\section{The Wigner function, the Krylov basis, and semiclassicality} \label{sec:prelim}
In this section, we review the formalism of Wigner negativity and the Krylov basis, which play central roles in our analysis, and introduce the notion of semiclassicality that we will develop throughout this work.
\subsection{The discrete Wigner function}
\label{sec:DiscWigner}
Any finite dimensional quantum system of odd, prime dimension $D$\footnote{The discrete Wigner function formalism is nicest when $D$ is either an odd prime  or a power of an odd prime, although most aspects of the formalism work more generally for odd $D$, and generalizations exist for even $D$ as well. When $D$ is an odd prime power, the Wigner function can be shown to be the unique quasi-probability distribution which ``geometrizes'' the action of the Clifford group in terms of symplectic affine transformations on phase space \cite{Gross_2006}. } has a natural quasi-probability representation called the \emph{discrete Wigner function} \cite{WOOTTERS19871, Leonard, sphere, quantumcomp, Galois, Gross_2006, classicality}. In Sec.~\ref{sec:InfiniteWigner}, we will generalize to a Wigner function for infinite dimensional systems ($D\to\infty$) for which the conditions on $D$ are absent.
 For now, let $\left\{|k\rangle\right\}_{k=0}^{D-1}$ be an ordered, orthonormal basis for the Hilbert space. The essential idea is to interpret this as the ``position'' basis, and to construct a corresponding ``phase space''.  Define the \emph{discrete phase space} $\mathcal{P}_D$ as a lattice $\mathbb{Z}_D \times \mathbb{Z}_D$ of size $D^2$. With respect to the chosen basis, we define a set of $D^2$ operators $A(q,p)$ called phase-point operators, each labeled by a phase space point $(q,p)$:
\begin{equation}\label{eq:A_Wooters}
A(q,p) = \sum_{k,\ell=0}^{D-1} \widehat{\delta}_{2q,k+l}e^{\frac{2\pi i}{D} (k-\ell)p }|k\rangle \langle \ell|,
\end{equation}
where $q$ and $p$ are integers and the hatted Kronecker delta $\widehat{\delta}$ is the $\text{mod}\,D$ version, i.e., it is 1 when $(k+\ell) = 2q\,\text{mod}\,D$, and 0 otherwise. The discrete Wigner function for a density matrix $\rho$ is now defined (in analogy with the continuous case \cite{Wigner}) as: 
\beq 
W_{\rho}(q,p) = \frac{1}{D} \mathrm{Tr}\left(\rho A(q,p)\right).
\eeq 
The Wigner function attempts to represent the quantum state $\rho$ as a probability distribution in phase space, much like in classical mechanics. Indeed, it satisfies the following properties:
\begin{enumerate}
\item The discrete Wigner function is real and unit normalized, i.e., $\sum_{q,p=0}^{D-1} W_{\rho}(q,p) = 1$, assuming $\text{Tr}\,(\rho)=1.$
\item Summing over one of the directions, say either $p$ or $q$, reduces the Wigner function to a probability density along the other direction:
\beq 
\sum_{p=0}^{D-1} W_{\rho}(q,p) = \langle q|\rho|q\rangle,\qquad\sum_{q=0}^{D-1} W_{\rho}(q,p) = \langle p|\rho|p\rangle,
\eeq 
where we have defined the ``momentum eigenstates'' as $|p\rangle = \frac{1}{\sqrt{D}}\sum_{q=0}^{D-1}\exp\left(\frac{2\pi i pq}{D}\right)|q\rangle$. 
\item The time evolution of the discrete Wigner function follows a discrete version of the Moyal equation \cite{WOOTTERS19871}, which can be thought of as the quantum analog of the Liouville equation from classical mechanics.  
\end{enumerate}
While the above properties seem to suggest that we should regard the Wigner function as a probability distribution in phase space, a simplistic interpretation of this kind fails for an important reason -- the Wigner function can take negative values at some points in phase space. That said, one might expect that states for which the Wigner function is everywhere non-negative are ``classical'' with respect to the chosen basis, and that  the ``amount'' of negativity in the Wigner function could be thought of as a measure of non-classicality. 

This expectation is made sharp by the \emph{Gottesman-Knill theorem} \cite{gottesman1998heisenberg, Aaronson:2004xuh}. In the theory of quantum computation, \emph{stabilizer circuits}  have a special property: they can be efficiently simulated on a classical computer.\footnote{Relatedly, any stabilizer/Clifford unitary can be synthesized from a polynomial number of elementary gates \cite{Hostens:2005svl}.} However, stabilizer circuits are not universal, and one needs additional ``magic'' gates to simulate arbitrary quantum circuits \cite{Bravyi_2005}. In the resource theory of magic \cite{Veitch_2014}, stabilizer operations are regarded as free, while magic or non-stabilizerness is regarded as a resource; see \cite{White:2020zoz, Cao:2023mzo, Cao:2024nrx, Basu:2025uxw, Malvimat:2026oqf, Bettaque:2026vpl} for some recent work on magic and stabilizer complexity in the gravity context. It turns out that the \emph{sum negativity} of the Wigner function, defined as:
\begin{equation}
    \label{negdef}
    \cN_s(\rho)= \frac{1}{2}\left(\sum_{q,p} |W_{\rho}(q,p)|-1\right),
\end{equation}
is a magic monotone, i.e., an operationally meaningful measure of non-stabilizerness \cite{Veitch_2014}. Since stabilizer circuits can be simulated efficiently on a classical computer, we can also interpret Wigner negativity as measuring the complexity of  simulating a quantum state and its dynamics on a classical computer by using the Wigner quasi-probability representation.\footnote{A word of caution: states with local tensor product structure and low entanglement can  have large Wigner negativity and yet be classically simulatable using other (i.e., non-quasi-probability) methods because of the low entanglement in the state \cite{Vidal:2003pmm}.}  In this paper, we will focus on the quantity
\beq \label{eq:conv}
\mathcal{N}(\rho) := 1+ 2\mathcal{N}_s(\rho)= \sum_{q,p}|W_{\rho}(q,p)|,
\eeq 
which we will call Wigner negativity. 

\subsection{Wigner negativity and and  semiclassicality}
\label{sec:semiclassical}
We  will treat Wigner negativity as a measure of semiclassicality with respect to a  choice of basis, and consider the degree to which an initially classical state remains semiclassical with respect to this measure as it evolves in time.  Noting that for a Hilbert space of dimension $D$ a typical state will have negativity $O(\sqrt{D})$ \cite{Basu:2025mmm}, we will say that a state is {\it classical} if it has no negativity, and {\it semiclassical} with respect to a given basis if its  negativity does not diverge as $D\to\infty$.   In this nomenclature states with diverging negativity in the limit of large system size are non-classical. Similarly, we will say that a {\it classical} dynamics does not increase negativity, and a  dynamics is {\it semiclassical} if the negativity potentially grows but remains $O(1)$ (i.e., does not scale with system size) under time evolution.  In the large system limit, a classical state experiencing semiclassical dynamics will retain $O(1)$ negativity after an $O(1)$ amount of time elapses.

The word semiclassical is often associated with the existence of a classical trajectory which dominates the path integral. In weakly-coupled quantum systems, there is usually a natural underlying basis (say, the position basis for a particle moving on a line under some potential) in terms of which the Hamiltonian is relatively simple. In such cases, it is natural to slice the path-integral by using this basis (and also its conjugate momentum basis), and if one or more classical trajectories dominate in the path integral,  we say that the time evolution is semiclassical. Equivalently, if one slices the path integral for a semiclassical time evolution at any moment of time,  the wavefunction at that time will be peaked around some classical trajectories with a small spread. Wigner negativity is intimately connected to this notion of semiclassicality -- indeed, the negativity of the discrete Wigner function with respect to some basis $\{|q\rangle\}_{q=0}^{D-1}$ gives a lower bound \cite{Basu:2025mmm}:
\beq 
\text{min}\left(H_{\frac{1}{2}}(q),H_{\frac{1}{2}}(p)\right)\geq \log \mathcal{N},
\eeq 
where $H_{\frac{1}{2}}(q)$ is the $1/2$-R\'enyi entropy of the probability distribution $P_q = |\langle q|\psi\rangle|^2$, and $H_{\frac{1}{2}}(p)$ is the $1/2$-R\'enyi entropy of the probability distribution in the conjugate momentum basis. 

In  large-$N$ systems with many degrees of freedom, there is a further subtlety. In such systems, as the number of degrees of freedom diverges, it is often helpful to ``change variables'' to some collective field description (see, for instance, \cite{Yaffe:1981vf, Jevicki, Das:1990kaa}), and the path integral is semiclassical in those degrees of freedom, not in the original elementary fields of the theory. Things get even less clear in strongly coupled systems where the Hamiltonian of interest is a complicated matrix (i.e., with some random matrix features) generating a chaotic time evolution. In this case, it is not always obvious how to find the correct variables that give a good semiclassical path integral description, should one exist. {\it A key goal of this paper is to use Wigner negativity together with Krylov space techniques (described below) to give a general construction of semiclassical variables for large many-body chaotic quantum systems.}
\subsection{The Krylov basis and complexity}
The notion of magic and negativity are defined with respect to a choice of basis. A primary goal of this work is to study the growth of these quantities under time evolution in the Krylov basis.  Given an initial state $\ket{\psi_0}$ and a Hamiltonian $H$, the Krylov basis is an orthonormal basis for the subspace of the Hilbert space explored by the time evolution of $\ket{\psi_0}$. More precisely, consider the set of states $\{\ket{\psi_0},H\ket{\psi_0}, H^2\ket{\psi_0},\cdots\}$, which span the so-called Krylov subspace. These states by themselves are not orthonormal, but we can construct an ordered orthonormal set of states $\{|K_n\rangle\}$, indexed by a {\it Krylov index}, by following the Lanczos algorithm (Gram-Schmidt procedure) applied to this subspace. By definition, in the Krylov basis, the Hamiltonian takes a tri-diagonal form:
\beq\label{tri}
H\ket{K_n}=a_n\ket{K_n}+b_n\ket{K_{n-1}}+b_{n+1}\ket{K_{n+1}},
\eeq
where the $a_n$ and $b_n$ are called the \emph{Lanczos coefficients}. Then the time-evolved state is expanded in the Krylov basis as
\be
\ket{\psi(t)}=e^{-iHt}\ket{\psi_0}=\sum^{d-1}_{n=0}\psi_n(t)\ket{K_n},
\ee
where $d$ is the dimension of the Krylov basis, which can be finite or infinite. Consequently, the wave functions $\psi_n(t)\equiv\langle K_n| e^{-itH} |\psi_0\rangle$ satisfy a discrete Schrodinger equation
\be
i\partial_t\psi_n(t)=a_n\psi_n(t)+b_n\psi_{n-1}(t)+b_{n+1}\psi_{n+1}(t),\qquad \psi_n(0)=\delta_{n,0}.\label{eq:SchrEq}
\ee

All the information about the Lanczos coefficients, and more generally, the wavefunctions $\psi_n(t)$, is contained in the return amplitude which is defined as:
\beqn
S(t)=\langle \psi(t)|\psi_0\rangle=\bra{\psi_0}e^{iHt}\ket{\psi_0}.
\eeqn
In particular, the Lanczos coefficients can be extracted from it using the moments:
\beq
\mu_n=\frac{d^n}{dt^n}S(t)=\bra{\psi_0}(iH)^n\ket{\psi_0}=\bra{K_0}(iH)^n\ket{K_0}.
\eeq
It follows that these moments are related in a specific way to the Lanczos coefficients; the first few of these relations are:
\beqn
\mu_0=1,\qquad\mu_1&=&-(a_0^2+b_1^2),\qquad
\mu_2=-i(a_0^3+2a_0b_1^2+a_1b_1^2),
\eeqn
and so on (see \cite{Balasubramanian:2022tpr}).  These methods are reviewed in \cite{Nandy:2024htc,Baiguera:2025dkc, Rabinovici:2025otw} (see also \cite{Paeckel:2019yjf} for review on the practical applications of these methods in time evolution of tensor network states).  There are also alternative analytical methods that derive the Lanczos spectrum directly as an integral formula \cite{Balasubramanian:2022dnj}, or in terms of orthogonal polynomials determined from the density of states \cite{Muck:2022xfc,Balasubramanian:2025xkj,Balasubramanian:2026azk}.  In general, given a Hamiltonian and initial state, it is hard to analytically solve for the Lanczos coefficients. Fortunately, there are  exactly solvable models (see Appendix \ref{App:solvable}) with known coefficients and we will utilize them below.

Once the Schrodinger equation has been solved and we know the probabilities $p_n(t)=|\psi_n(t)|^2$, the spread/Krylov complexity is defined as \cite{Parker:2018yvk,Balasubramanian:2022tpr}
\be
C(t)=\sum^{d-1}_{n=0}n|\psi_n(t)|^2\,.\label{SpreadK}
\ee
A related quantity is the Krylov entropy \cite{Barbon:2019wsy} given by the Shannon entropy of the probability distribution $p_n(t)$:
\be
S_K(t)=-\sum^{d-1}_{n=0}p_n(t)
\log(p_n(t))\,.
\ee
The quantity $e^{S_K}$ can be regarded as a measure of the dimension of the subspace within which the time-evolved state is substantially spread over in the Krylov basis \cite{Balasubramanian:2022tpr}. We will contrast it below with spread complexity and Wigner negativity in various solvable models.  

\subsection{Infinite dimensional Krylov-Wigner function}
\label{sec:InfiniteWigner}
In the following, we will study the Wigner negativity with respect to the Krylov basis of a time evolved pure state:
\beq 
|\psi(t)\rangle = e^{-itH} |0\rangle,
\label{eq:samplestate}
\eeq 
where $|0\rangle$ is the specified initial state (not necessarily the vacuum). In most of the cases we will encounter, the Krylov basis will be infinite dimensional.  Thus we will have to extend the finite dimensional formalism in Sec.~\ref{sec:DiscWigner} to infinite dimensional Hilbert spaces, and thereby  define a {\it Krylov-Wigner function} associated to time evolution in the Krylov basis. To this end, we first construct a family of discrete Wigner functions by truncating the Krylov chain at some large (prime) value $D$ of the Krylov index, and  then take the  $D\to \infty$ limit. One nice feature of time evolution in the Krylov basis is that the wavefunction spreads in a local manner along the Krylov chain because the tridiagonal form of the Hamiltonian implies local hopping dynamics. This  means that if one is interested in dynamics for a large but $O(1)$ time $t$, then one can  truncate the Hilbert space at a very large value $D \gg t$ of the Krylov index. At times much smaller than $D$, the projection onto the truncated Hilbert space does not have a significant effect on the time evolved state $\ket{\psi(t)}$ (see, for instance, theorem 2 in \cite{Hochbruck} for systems with bounded energy support). Thus, we expect that we can define a Krylov-Wigner function within the truncated Hilbert space, and define the infinite-dimensional version by taking the $D\to \infty$ limit. This approach has been used previously in \cite{Barnett, ALuk1993, HCH} (see also \cite{Mukunda, Bizarro} for related older work) to define ``number-phase Wigner functions'' in the context of quantum optics. 

Let us start with the Krylov-Wigner function for the state (\ref{eq:samplestate}) in the truncated Hilbert space
\beq \label{KWan}
W_{(D)}(q,p) = \frac{1}{D}\sum_{k,\ell=0}^{D-1}\widehat{\delta}_{2q,k+\ell}e^{\frac{2\pi i}{D}(k-\ell)p} \langle k|e^{-itH}|0\rangle \langle 0| e^{itH}|\ell\rangle,
\eeq
where $|k\rangle$ and $|\ell\rangle$ are Krylov basis vectors, $q$ and $p$ are integers, $D$ is prime, and the delta function $\widehat{\delta}$ imposes the condition $2q = k+\ell \mod D$.   For $O(1)$ values of time such that $t \ll D$, the amplitude $\langle k | e^{-itH}|0\rangle$ is essentially contained in the region where $k$ is $O(1)$ while $\frac{k}{D} \to 0$ as $D\to\infty$, so the above projection onto the truncated Hilbert space is innocuous. 

Now, consider for a moment perturbatively expanding the Wigner function around $t=0$. In the Krylov basis the state starts out localized on $|0\rangle$ at $t=0$, and then spreads systematically to $|1\rangle, |2\rangle$ etc. At linear order in $t$, the state is localized on $|0\rangle$ and $|1\rangle$, and so in the Wigner function, the terms $k=1,\ell =0$ or $\ell=0,k=1$ contribute. In both cases the solution to the delta function is $q=\frac{D+1}{2}$. More generally, what happens is that as the wavefunction in Krylov space spreads locally on the Krylov chain $0 \to 1\to 2$ etc., the Wigner function spreads in $q$ from $0 \to \frac{D+1}{2} \to 1 \to \frac{D+3}{2}$ and so on.  This is awkward, since we want to take $D\to\infty$ limit. But the resolution is straightforward; in order to take $D\to \infty$, we instead take $q$ to be a half-integer ranging over $q=0,\frac{1}{2},1,\cdots, \frac{D-1}{2}$.  The point is that both $q=1/2$ or $q=(D+1)/2$ give the same value of $2q\mod D$, and thus, we can restrict the range of $q$ to half-integers $0,1/2,\cdots (D-1)/2$.  We emphasize that this is just a re-labeling of the entries of the Wigner function to facilitate the $D\to\infty$ limit.

Additionally, at early $O(1)$ times when $k$ and $\ell$ are both $O(1)$, we can replace the modular delta function by the regular Kronecker delta since $k+ \ell$ will be much smaller than $D$ anyway. Finally, if we define the ratio
\beq 
\theta = \frac{2\pi p}{D},
\eeq 
we can treat the momentum as a continuous periodic variable as $D$ becomes large. Presented in this way, the effective phase space has a discrete position variable ranging over non-negative, half-integer values and a continuous, periodic momentum variable. Defining 
\beq 
W_{(D)}(q,p) = \frac{2\pi}{D} W(q,\theta = \frac{2\pi p}{D}),
\eeq 
we get the infinite-dimensional Krylov-Wigner function:
\beq \label{WigKry} 
W(q,\theta) = \frac{1}{2\pi}\sum_{k,\ell=0}^{\infty}\delta_{2q,k+\ell}e^{i(k-\ell)\theta} \psi_k(t) \psi_{\ell}^*(t).
\eeq 
The negativity of this Wigner function is given by:
\beq 
\mathcal{N}(\psi_t) = \int \frac{d\theta}{2\pi} \sum_{q=0,\frac{1}{2},1\cdots} |W(q,\theta)|,\label{eq:negdef}
\eeq 
which is simply the $D\to \infty$ limit of the negativity of the discrete Wigner function.

In \cite{Mari_2012}, it was shown in the finite $D$ case that for quantum circuits involving stabilizer gates, the outcome of the circuit can be efficiently estimated on a classical computer. This is possible because a stabilizer state becomes a classical probability distribution via its Wigner function, while stabilizer gates become classical stochastic matrices via their Wigner function representations. The outcome of the quantum circuit can then be estimated by Monte Carlo sampling from classical trajectories using the Wigner function representations of individual circuit elements. In \cite{Pashayan_2015} this analysis was extended to circuits involving non-stabilizer elements, and it was shown that the cost of similarly estimating the outcome of a quantum circuit is proportional to the total amount of Wigner negativity in the circuit.  It would be interesting to use the methods in \cite{Pashayan_2015} to similarly show that Wigner negativity in the Krylov basis is related to computational cost even in the $D\to \infty$ limit.   We expect all the steps in the arguments of \cite{Pashayan_2015} to go through,  giving an $O(1)$ cost at any $O(1)$ time.\footnote{The estimation cost may scale exponentially with time, but will be independent of the Hilbert space dimension.} It would also be interesting to see if one could construct a resource theory around the negativity of this infinite-dimensional Krylov-Wigner function by identifying stabilizer protocols which admit a good $D\to \infty$ limit.  We will leave these interesting questions for the future.

\section{Wigner negativity in 2d CFT} \label{sec:2dCFT}
In this section, we will study the time evolution of Wigner negativity for excited states of a 2d CFT, obtained by acting with a light operator with $O(1)$ dimension on either the vacuum or TFD state. These are examples of local operator quenches in CFTs \cite{Nozaki:2014hna} and their spread complexity was studied in \cite{Caputa:2023vyr,Caputa:2025dep}. Moreover, the relation between the rate of growth (time derivative) of the spread complexity for these states and proper radial momentum of massive particles in $AdS_3$ was established in \cite{Caputa:2024sux}. This correspondence suggests an emergent classical picture for dynamics in these states which we will substantiate below with precise results on Wigner negativity. 

\subsection{Three examples for studying negativity growth}
\label{sec:threesettings}

We will consider three different settings in $\text{CFT}_2$ to study negativity growth. We note that in all these cases, the Hilbert space dimension is effectively infinite to begin with, and so we are always in a regime where $t$ is $O(1)$. 

\subsubsection*{Example 1: Vacuum state on an infinite line}
Consider a $2d$ CFT  on an infinite line. Let the initial state $\ket{\psi_0}$ to be of the form of a local operator quench \cite{Nozaki:2014hna}:
\beq
\ket{\psi_0}=\mathcal{N}_{\epsilon}\,e^{-\epsilon H} \mathcal{O}(x_0)\ket{0},
\eeq
where $|0\rangle$ is the vacuum  and $\mathcal{O}(x_0)$ is a primary operator of conformal dimension $\Delta$ inserted at spatial position $x_0$ on the line. The evolution over Euclidean time $\tau=\epsilon$ renders the state normalizable, with finite energy $E\sim \Delta/\epsilon$. The  factor $\mathcal{N}_{\epsilon}$ is chosen to make the state $\ket{\psi_0}$ have unit norm. In a CFT with a holographic dual, if we further take $\mathcal{O}$ to be a light, single-trace primary, then in the bulk this state corresponds to a particle in the Poincare patch of empty $\text{AdS}_3$ with  mass  $(mL_{AdS})^2=-\Delta(d-\Delta)$. As we evolve this state in time, the particle moves on a timelike geodesic, falling inwards \cite{Nozaki:2013wia}. 

In order to calculate the wavefunction in the Krylov basis, we first calculate the return amplitude, which can be extracted from the vacuum $2-$point function of the operator $\mathcal{O}$:
\beq
S(t)=\left(\frac{2i\epsilon}{t+2i\epsilon}\right)^{2\Delta}.
\eeq
This matches the $\text{SL}(2,\mathbb{R})$ return amplitude for the highest weight representation $h=\Delta$\footnote{Representation weight $h$ should not be confused with the chiral operator dimension in CFT $\Delta=h_{CFT}+\bar{h}_{CFT}$.} given in Eq.~\ref{eq:return} of Appendix~\ref{App:SL2R}, if we set the parameters in that expression  as
\beqn
\alpha=\frac{1}{2\epsilon}, \ \ \ \gamma=\frac{1}{\epsilon}, \ \ \ \delta=0, \ \ \ \mathcal{D}=0 \, .
\eeqn
Hence we can obtain the Lanczos coefficients and the Krylov wavefunction analytically from the general  $\text{SL}(2,\mathbb{R})$ formulae.

\subsubsection*{Example 2: TFD state on an infinite line}
Consider the initial state:
\beq
\ket{\psi_0}=\mathcal{N}_{\epsilon} e^{-\epsilon H_R} \mathcal{O}_R(x_0)e^{\epsilon H_R} e^{-\frac{\beta}{2} H_R}\ket{\Omega},
\eeq
where $\ket{\Omega}$ is the maximally mixed state between two copies (left and right) of the CFT, and we  think of the operator and the Euclidean Hamiltonian evolution as acting on the right factor. Holographically, this corresponds to a particle in the right exterior of the planar BTZ black hole geometry. In the TFD case, we can consider two different time evolutions for such an excited state: (a) with the relative Hamiltonian $(H_R - H_L)$, or (b) with the total Hamiltonian $(H_L + H_R)$, or equivalently with the one sided Hamiltonian $H_R$. For case (a), the return amplitude is given by:
\beq
S(t) = \left( \frac{\text{sinh}\left(\frac{\pi (t+2 i \epsilon)}{\beta} \right)}{\text{sinh}\left( \frac{2\pi i \epsilon}{\beta} \right)} \right)^{-2\Delta},
\eeq 
and once again matches the $SL(2,\mathbb{R})$ return amplitude with $h=\Delta$ (see Appendix~\ref{App:SL2R}) if we set the parameters in Eq.~\ref{eq:return} as
\beq 
\alpha = \frac{\pi}{\beta \ \text{sin}\left(\frac{2 \pi \epsilon}{\beta} \right)}, \ \ \ \gamma = \frac{2\pi}{\beta} \text{cot}\left(\frac{2 \pi \epsilon}{\beta}\right), \ \ \ \delta=0, \ \ \ \mathcal{D}=\frac{2\pi}{\beta}.
\eeq
For case (b), the return amplitude is given by:
\be
S(t)=
\left(\frac{(\beta-2it)\sinh\left(\frac{2\pi i\epsilon}{\beta-2it}\left(1-\frac{it}{2\epsilon}\right)\right)}{\beta\sinh\left(\frac{2\pi i\epsilon}{\beta}\right)}\right)^{-2\Delta}.\label{RetAmplHLHR}
\ee
To our knowledge, there is no known closed form expression for the Lanczos coefficients and the Krylov wavefunctions in this case. In the rest of this work, we will only consider case (a), where analytic progress can be made.  

\subsubsection*{Example 3: Vacuum state on a circle}
In this case, the initial state is just as in case 1, except that the  CFT is  on a circle of length $L$. The corresponding return amplitude is:
\beq
S(t) = \left( \frac{\text{sin}\left(\frac{\pi (t+2 i \epsilon)}{L} \right)}{\text{sin}\left( \frac{2\pi i \epsilon}{L} \right)} \right)^{-2\Delta},
\eeq 
and once again takes the $SL(2,\mathbb{R})$ form from Appendix~\ref{App:SL2R}, with $h=\Delta$ and
\beq 
\alpha = \frac{\pi}{L \ \text{sinh}\left(\frac{2 \pi \epsilon}{L} \right)}, \ \ \ \gamma = \frac{2\pi}{L} \text{coth}\left(\frac{2 \pi \epsilon}{L}\right), \ \ \ \delta=0, \ \ \ \mathcal{D}=\frac{2\pi i}{L}.
\eeq
\subsection{Wigner negativity}
As described above, in the three examples we study the Krylov dynamics are solvable in terms of the $SL(2,\mathbb{R})$ model;  we will use these solutions to study the Wigner function and its negativity. Recall that the (infinite-dimensional) Krylov-Wigner function is given by:
\beq 
W(q,\theta) = \frac{1}{2\pi}\sum_{m,n=0}^{\infty}e^{i\theta(m-n)}\delta_{2q,m+n}\psi_m(t) \psi^*_n(t) ,\label{eq:wig}
\eeq
where $q=0, {1 \over 2},1, \cdots, {D-1 \over 2}$, and $\theta$ is a momentum variable with $2\pi$ periodicity. Using the explicit form of the wavefunction in the $SL(2,\mathbb{R})$ model (see Eq.~\ref{eq:wvfnc}),
\beq 
\psi_n(t) = e^{-i \delta} \sqrt{\frac{\Gamma(2\Delta +n)}{\Gamma(2\Delta)\Gamma(n+1)}} g^{2\Delta} f^n \,,
\label{eq:wavefn}
\eeq 
where $f$ and $g$ are functions of time defined as:
\beqn
f = \frac{-2 i \alpha}{\mathcal{D}} \frac{\text{tanh}\left(\frac{\mathcal{D} t}{2}\right)}{1+\frac{i \gamma}{\mathcal{D}}\text{tanh}\left(\frac{\mathcal{D} t}{2}\right)},\qquad g = \frac{1}{\cosh(\frac{\mathcal{D} t}{2})+\frac{i \gamma}{\mathcal{D}}\sinh(\frac{\mathcal{D} t}{2})} \,,
\label{eq:fandg}
\eeqn
with $\mathcal{D}=\sqrt{4 \alpha^2-\gamma^2}$, we get
\beq \label{WigKry2}
W(q,\theta) = \frac{1}{2\pi\Gamma(2\Delta)}|g|^{4 \Delta} |f|^{2 q}\sum_{m=0}^{2 q}e^{i(\theta+\phi)(2m-2q)}  \chi(m;2q) \, .
\eeq
Here we have written $f=|f|e^{i\phi}$, and
\beq 
\chi(m;2q) = \sqrt{\frac{\Gamma(2\Delta+m) \Gamma(2\Delta+2q-m)}{ \Gamma(2q-m+1) \Gamma(m+1)}}. 
\eeq 
Note that the phase of $g$ and the phase $e^{-i\delta}$ cancel out in the Wigner function because these are independent of the Krylov index.
The parameters $\alpha$ and $\gamma$ are time-independent parameters that are functions of fixed quantities like the temperature and the size of the circle on which the theory is defined. We will use this result below for numerics as well as for analytical scaling arguments.

\subsubsection{Numerical results}
\begin{figure}[t]
 \centering
 \begin{tabular}{c c}
 \includegraphics[width=0.45\textwidth]{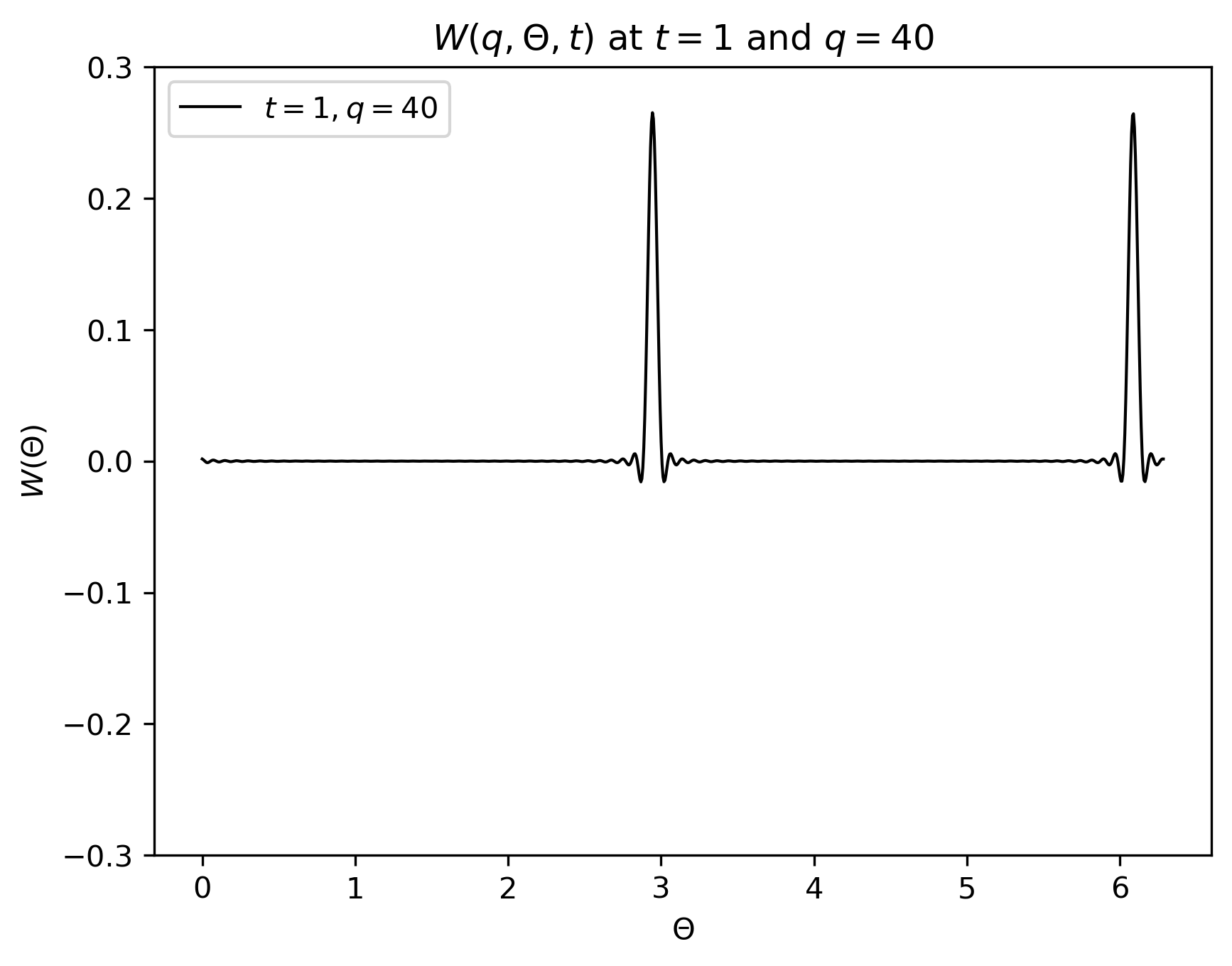} & \includegraphics[width=0.45\textwidth]{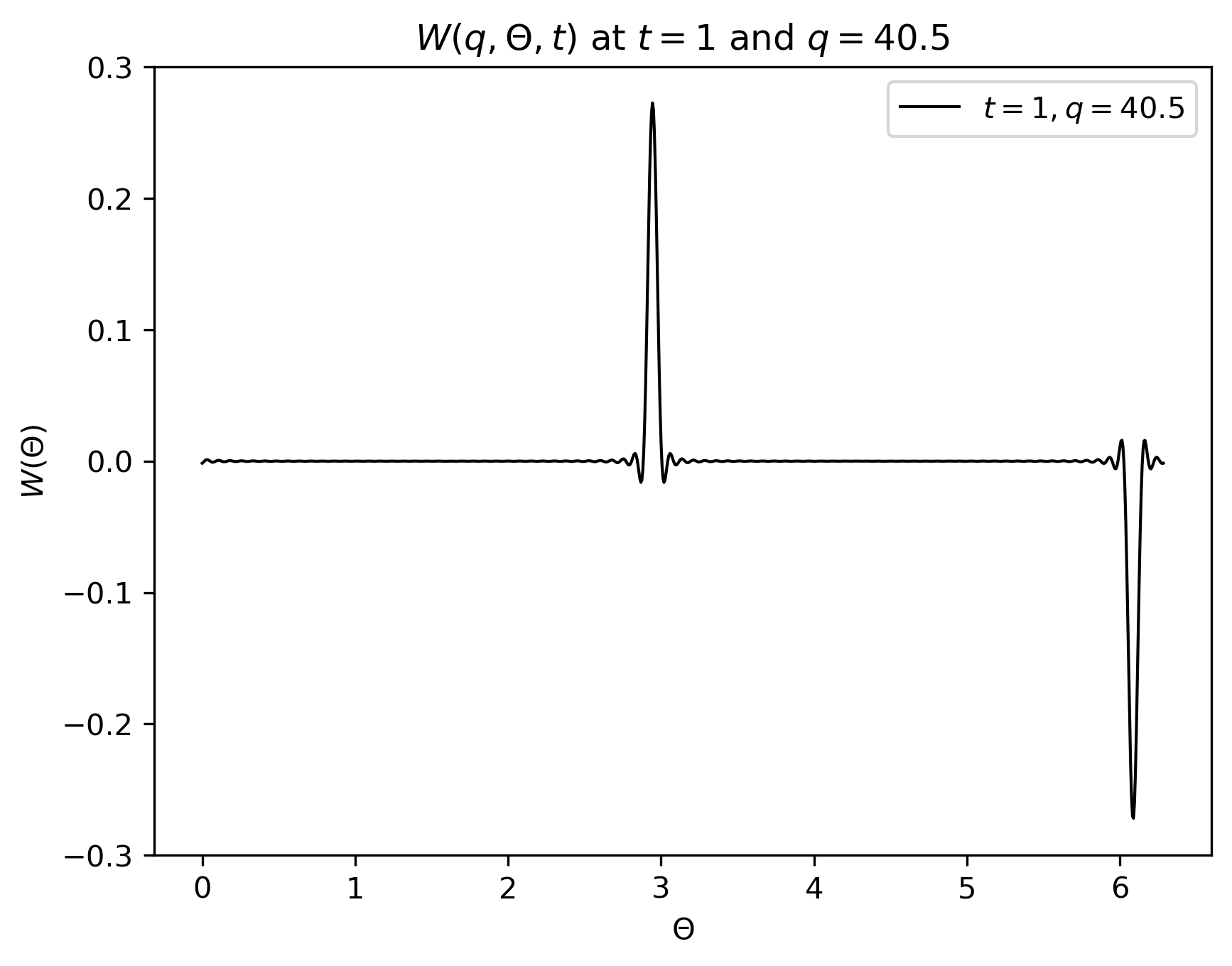}
 \end{tabular}
    \caption{The Wigner function $W(q,\theta)$ for the evolving vacuum state of a 2d CFT, perturbed by a primary operator of dimension $\Delta=2$, on an infinite line.  Results shown for integer $q$ at $t=1$, $q=40$ (left) and half-integer $q$ at $t=1$ and $q=40.5$ (right) as a function of $\theta$. The regularization parameter is $\ep=0.1$.}
    \label{fig:wig_SL2R_qfix}
\end{figure}

Fig.~\ref{fig:wig_SL2R_qfix} shows the  Wigner function as function of $\theta$ for fixed integer and half-integer $q$ for example 1 (the vacuum on an infinite line).  For integer $q$, the Wigner function has two peaks, close to $\theta=0,\pi$ (the peak near $0$ appears close to $2\pi$ since $\theta$ is a periodic variable; also the peak is not exactly at $0,\pi$ but approaches these values at large time.) For half integer $q$ it has a peak near $\theta=\pi$ and an anti-peak near $\theta=0$. We will analytically explain the origin of these features for $t\gg\ep$ in Sec.~\ref{sec:scaling}.  These features will become important below in understanding the behavior of Wigner negativity. 

\begin{figure}[t]
 \centering
 \begin{tabular}{c c}
 \includegraphics[width=0.45\textwidth]{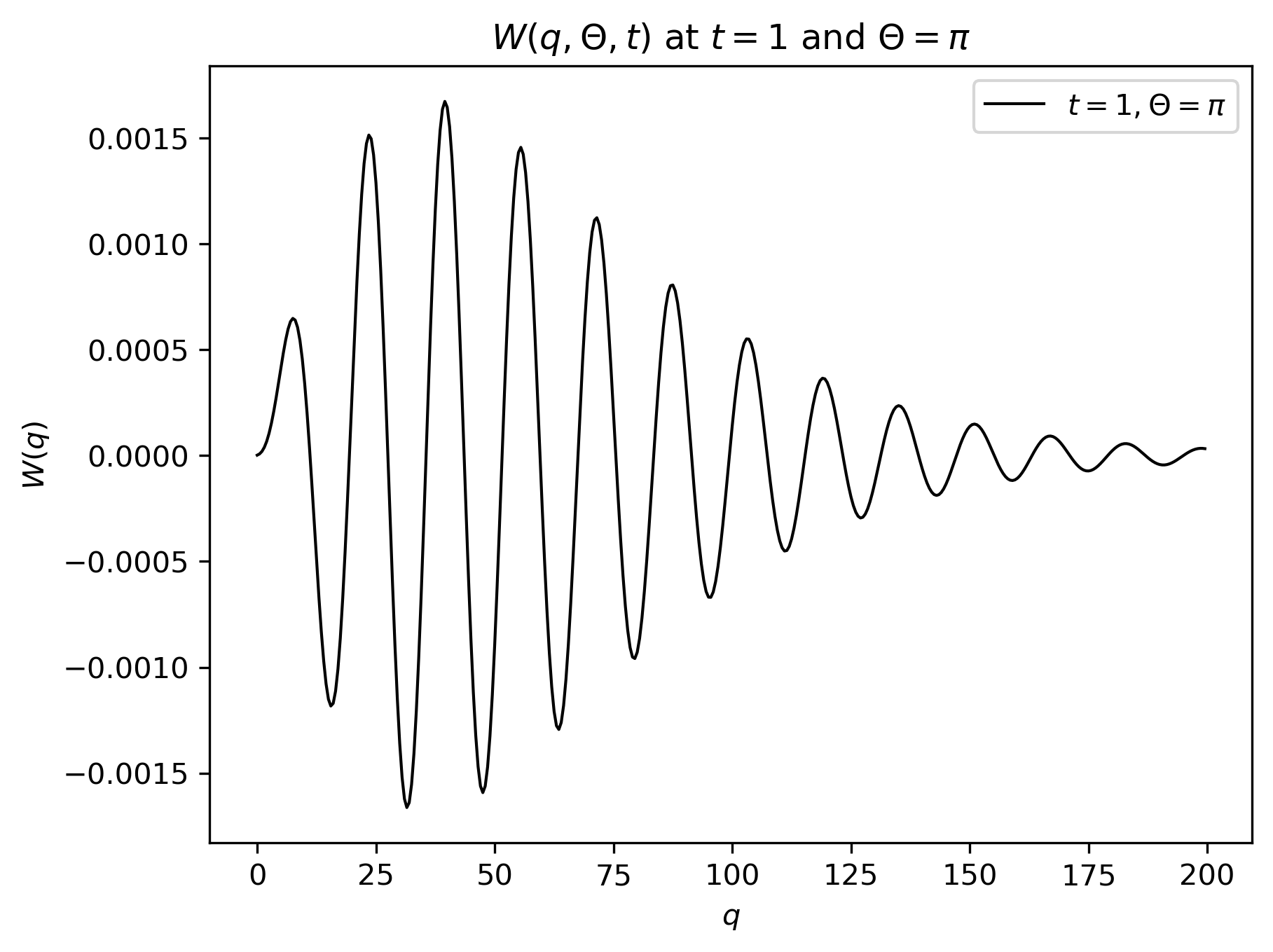} & \includegraphics[width=0.45\textwidth]{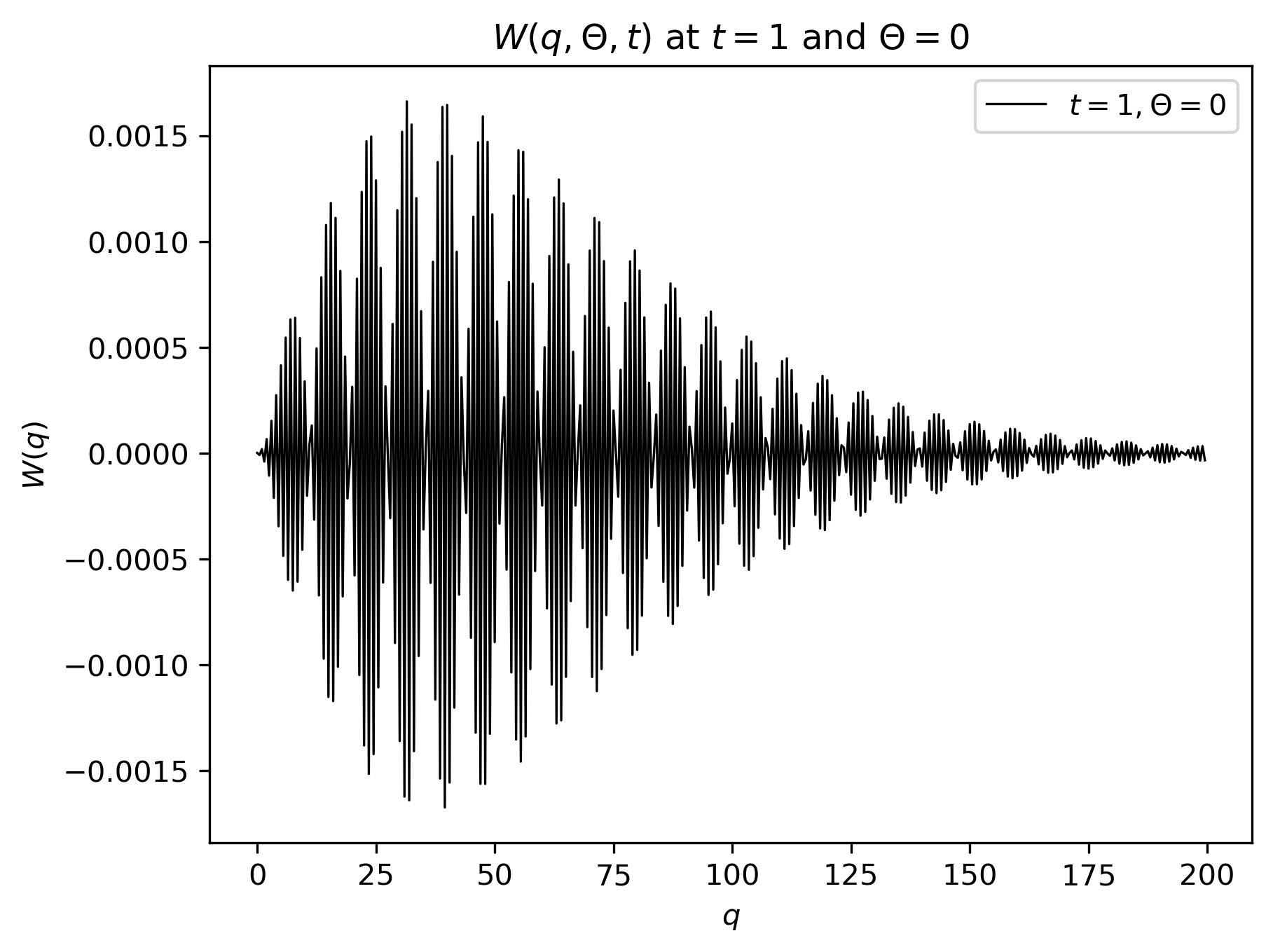}
 \end{tabular}
    \caption{ The Wigner function $W(q,\theta)$ for the evolving vacuum state of a 2d CFT, perturbed by a primary operator of dimension $\Delta=2$, on an infinite line. Results shown at $t=1$, $\theta=\pi$ (left) and $\theta=0$ (right) as a function of $q$. The regularization parameter is $\ep=0.1$. }  
    \label{fig:wig_SL2R}
\end{figure}

Fig.~\ref{fig:wig_SL2R} shows the  Wigner function as function of $q$ for fixed $\theta = \pi,0$ (the figure interpolates between the values obtained for $q=0,1/2,1,\cdots$). The oscillating curves are of a much smaller magnitude than Fig.~\ref{fig:wig_SL2R_qfix} because the sharp peak in the latter is not precisely at $\theta = \pi$; hence the value sampled at $\theta = \pi$ is much smaller, and oscillates for different $q$. The Wigner function depends on $q$ more smoothly when  $\theta=\pi$ (left panel of Fig.~\ref{fig:wig_SL2R}), since from Fig.~\ref{fig:wig_SL2R_qfix} it has a peak near this value of theta for both integer and half-integer $q$. The Wigner function is highly oscillatory at $\theta=0$ because,  from Fig.~\ref{fig:wig_SL2R_qfix}, near this value of $\theta$ the Wigner function has a peak for integer $q$ and an anti-peak for half-integer $q$.   The Wigner function in Example 2 (TFD state on an infinite line) behaves similarly for times $t \gg \beta$ and in Example 3 (vacuum state on a circle) for times $L \gg t \gg \epsilon$.

\begin{figure}[t]
    \centering
    \includegraphics[width=15cm, height=5.5cm]{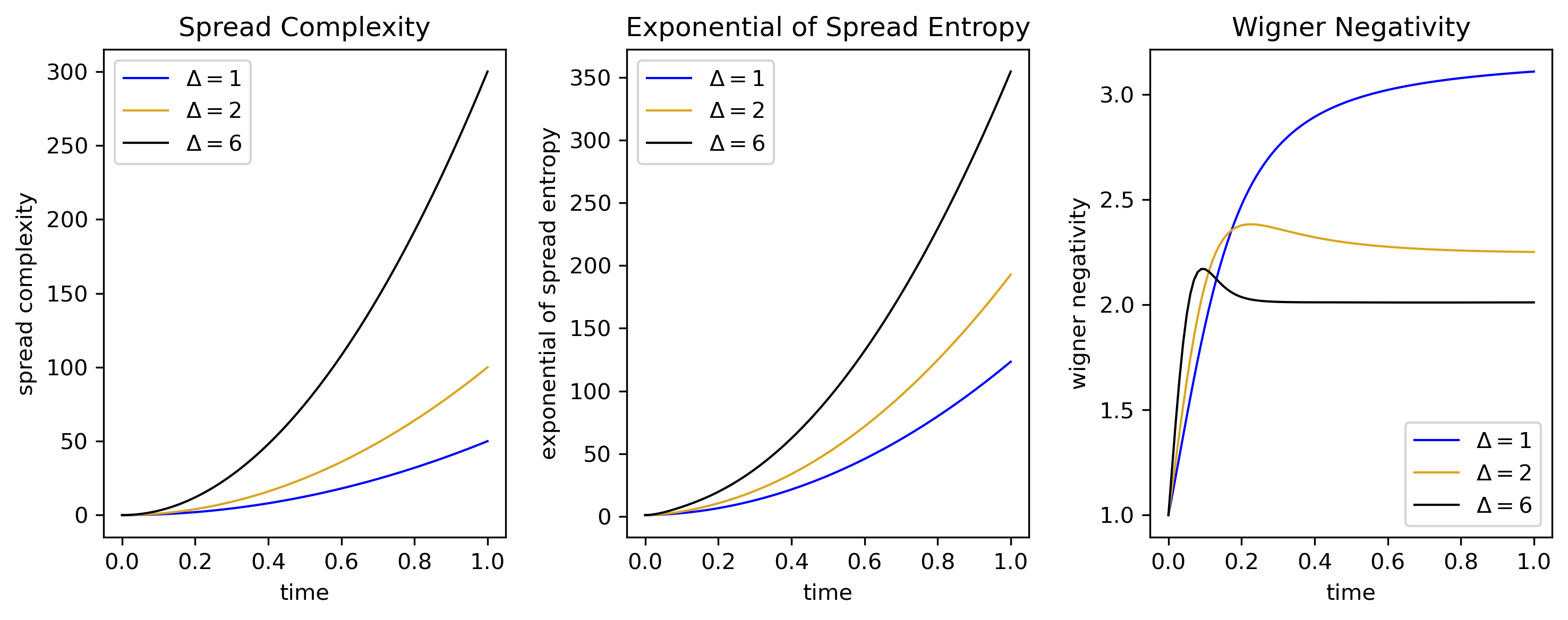}
    \caption{A comparison between spread complexity, exponential of spread entropy and Wigner negativity as a function of time for the  evolving vacuum state of a 2d CFT on an infinite line, perturbed by primary operators of dimension $\Delta=1,2,6$ and $\epsilon=0.1$.  The Wigner negativity saturates to an $O(1)$ constant while the spread complexity and spread entropy grow rapidly in time.  When $h=1$ the negativity saturates at times longer than those shown.
    }
    \label{fig:case_1}
\end{figure}

\begin{figure}[t]
    \centering
    \includegraphics[width=15cm, height=5.5cm]{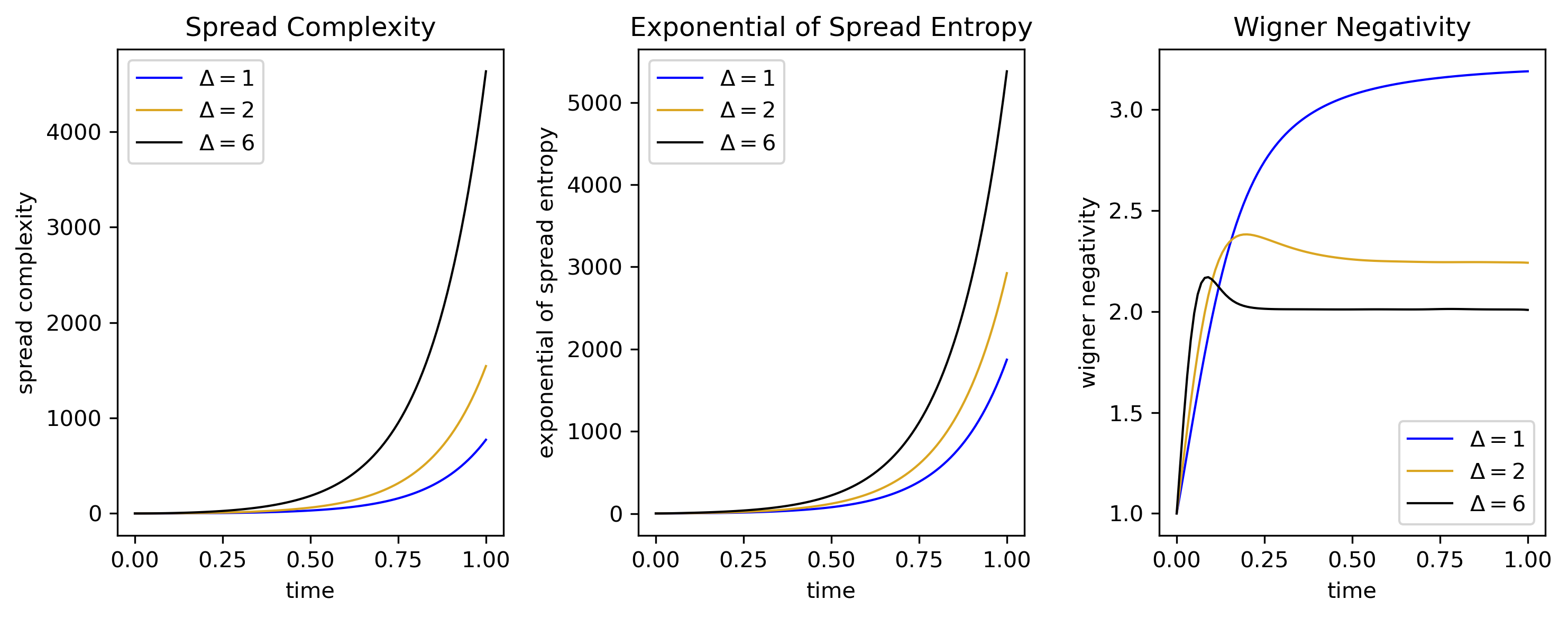}
    \caption{A comparison between spread complexity, exponential of spread entropy and Wigner negativity as a function of time for an evolving TFD state of a CFT on an infinite line, perturbed by primary operators of dimension  $\Delta=1,2,6$.  Here $\ep=0.1$ and $\beta=1$. The Wigner negativity saturates to an $O(1)$ constant while the spread complexity and and entropy grow rapidly in time.  When $h=1$ the negativity saturates at times longer than those shown.
    }
    \label{fig:case_2}
\end{figure}

\begin{figure}[t]
    \centering
    \includegraphics[width=15cm, height=5.5cm]{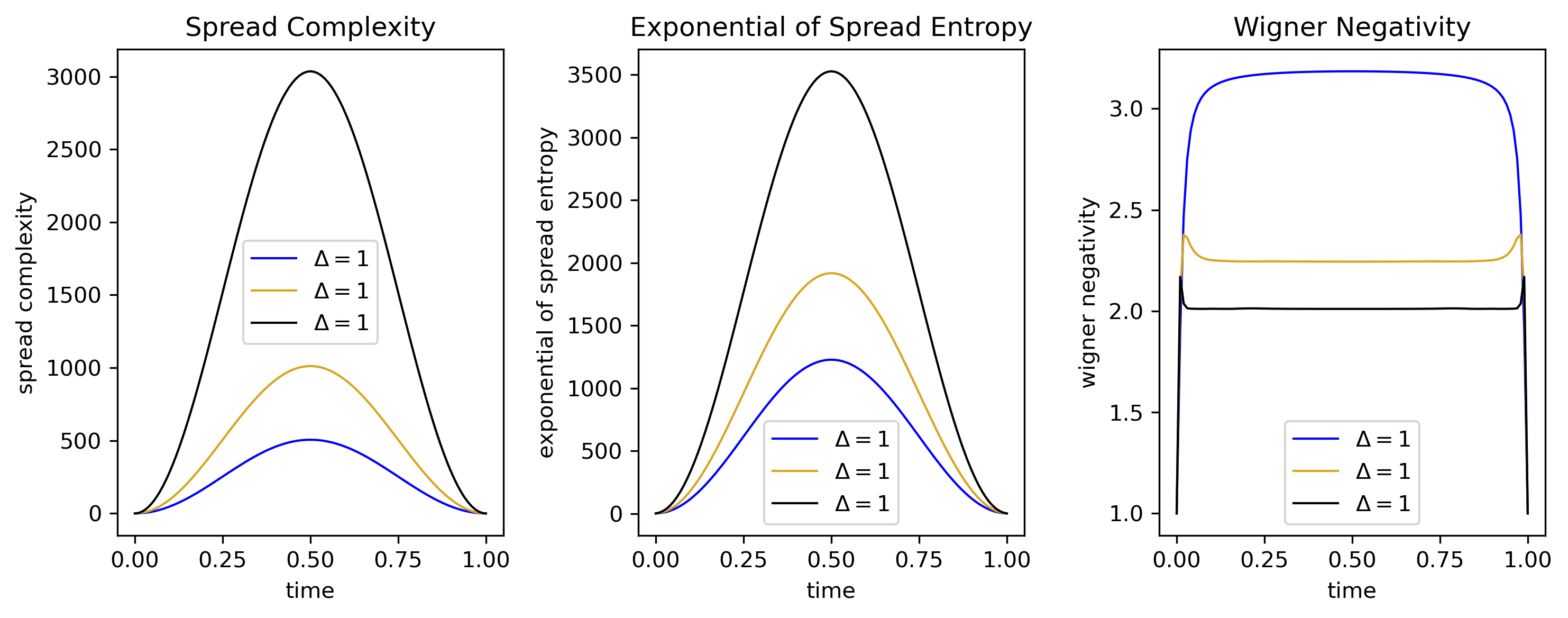}
    \caption{Comparison between spread complexity, exponential of spread entropy and Wigner negativity as a function of time for an evolving vacuum of a CFT on a circle of length $L$, perturbed by primary operators of dimension $\Delta=1,2,6$.  Here $\ep=0.1$ and $L=1$. There is a plateau for the Wigner negativity in the duration $\ep \ll t \ll L-\ep$, while the spread complexity and entropy increase to a peak and then decline again.  The decline to zero occurs because of the periodicity of the system on a circle. 
    }  
    \label{fig:case_3}
\end{figure}

Figs.~\ref{fig:case_1}, \ref{fig:case_2} and \ref{fig:case_3} show  the Wigner negativity, spread complexity, and spread entropy computed numerically for the three examples laid out in Sec.~\ref{sec:threesettings}. In all three cases, negativity starts growing initially, and then quickly saturates to an $O(1)$ value; the saturation happens for $t \gg \epsilon$ in the vacuum cases, and $t \gg \beta$ in the thermal case.  In sec. \ref{sec:scaling} we will explain why this saturation occurs, and  will analytically compute the saturation value.  By contrast, the  spread complexity has a quadratic, i.e., power-law, growth in time in example 1 (i.e., operator excitation on top of the vacuum on the line), while in example 2 (i.e., in the thermal case), it grows exponentially with time \cite{Balasubramanian:2022tpr,Caputa:2024sux}. This suggests that, while the wavefunction is spreading rapidly in the Krylov basis (as illustrated by the growth of the spread complexity), the complexity of classical simulation remains $O(1)$, and so the system actually remains approximately \emph{classical}. 
Fig.~\ref{fig:case_3} shows an additional feature -- for the CFT on a circle, at times of $O(L)$ (i.e., the length of the circle) the spread complexity and the negativity both decline to zero again, and show periodic behavior with the period equal to $L$.

A direct way to understand why this happens is to consider the wavefunction in momentum space $\widetilde{\psi}(\theta,t)$:
\beq
\widetilde{\psi}(\theta,t)=\sum_{k=0}^{\infty} e^{i\theta k} \psi_k(t)\,.
\eeq
Fig.~\ref{fig:mom SL(2,R) h=2} shows  $|\widetilde{\psi}(\theta,t)|^2$ for the first example (excitation around the vacuum) discussed above; the story for other cases is  similar (not shown). We  see that even as the wavefunction spreads in the Krylov basis, it gets squeezed and becomes peaked in the conjugate momentum basis around $\theta=\pi$. In this sense, the state becomes classical in momentum space, and this feature is captured by the Wigner negativity.

\begin{figure}[t]
    \centering
    \includegraphics[width=15cm, height=5.5cm]{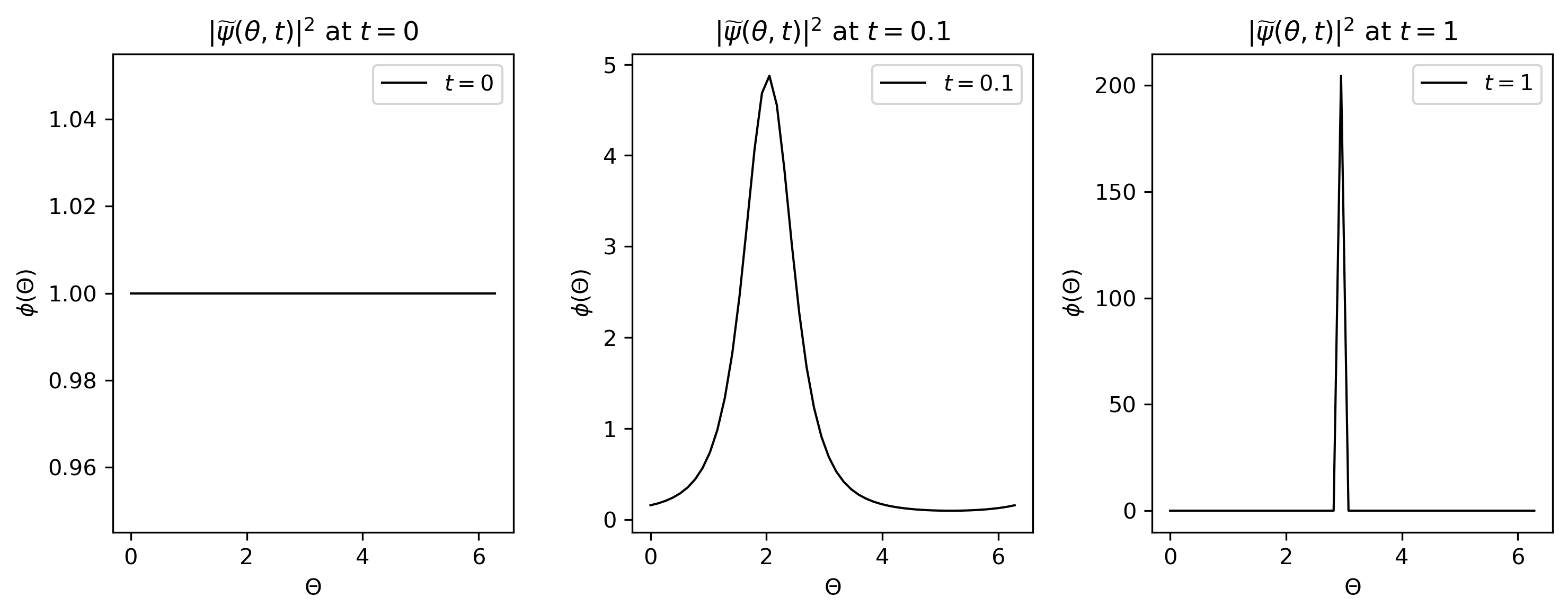}
    \caption{$|\widetilde{\psi}(\theta,t)|^2$ as a function of $\theta$ at 3 different times, i.e., $t=0$, $t=0.1$ and $t=1$ for $\Delta=2$ at $\ep=0.1$. We see that the wavefunction gets peaked in momentum basis at $\theta = \pi$ as time progresses.} \label{fig:mom SL(2,R) h=2}
\end{figure}

\subsubsection{Analytical results in a scaling limit} \label{sec:scaling}
We can understand the above results analytically at late times by considering the limiting forms of the functions $f$ and $g$ (\ref{eq:fandg}) that control the time dependence of the Wigner function (\ref{WigKry2}).  We will first extract these limiting forms for all three examples discussed above, and then use them to study the Wigner function.   Note that for all the cases that we consider, the return amplitude is controlled by a universal, kinematic CFT two-point, and finite-$N$ effects at late times are invisible. In this sense all finite times, even large ones, are  $O(1)$.

\noindent\textbf{Example 1 -- Perturbed CFT vacuum on a line}: When $t \gg \epsilon$,  we can expand $f$ and $g$ in (\ref{eq:fandg}) as functions of $t$ and write:
\beqn
|g| &=& \frac{2 \epsilon}{t}, \\
|f| &=& 1- \frac{2 \epsilon^2}{t^2}.
\eeqn
So we see that $g \to 0$ and $f \to 1$ when $t\gg\ep$.

\noindent\textbf{Example 2a -- Perturbed TFD state on a line}: In this case, when $t \gg \beta$, we can expand $f$ and $g$ as:
\beqn
|g|&=& 2 \ \text{sin}\left(\frac{2 \pi \ep}{\beta}\right) e^{-\frac{\pi t}{\beta}}, \\
|f| &=& 1-2 \ \text{sin}^2\left(\frac{2 \pi \ep}{\beta}\right) e^{-\frac{2 \pi t}{\beta}}.
\eeqn 
Again,  when $t\gg\beta$, $g \to 0$ and $f \to 1$. 

\noindent\textbf{Example 3 -- Perturbed CFT vacuum on a circle}
In this case we should  consider a limit in which the time is large, but also smaller than the scale set by the circle size, at which finite size effects will matter.  So consider times $t$ such that $\ep \ll L$ and $\ep \ll t \ll L-\ep$. In this regime, we find
\beqn
|g|&=&\frac{2 \pi \ep}{L \ \text{sin}\left( \frac{\pi t}{L}\right)},\\
|f| &=& 1-\frac{2 \pi^2 \ep^2}{L^2 \ \text{sin}^2\left(\frac{\pi t}{L} \right)}.
\eeqn

 The expressions above show that $|f|$ is infinitesimally close to $1$ at late times in the first two examples, and at large times smaller than the circle size in the third example.
Let us write $|f|=1-\eta(t)$, where $\eta(t)$ is infinitesimally close to $0$.   We then note that at large times, the support of the wavefunction spreads to larger values of $n$.  At such larger $n$ the Gamma functions in (\ref{eq:wavefn}) can be approximated by Stirling's formula giving

\beq 
\psi_n(t) \sim \frac{1}{\sqrt{\Gamma(2\Delta)}}[g(t)]^{2\Delta}\,n^{\Delta -\frac{1}{2}} (1-\eta(t))^n\,e^{in\phi(t)}.
\eeq 
Examining the saddle point of this expression, we see that the wavefunction on the Krylov chain at late times is peaked at $n_* \sim \frac{(2\Delta-1)}{\eta(t)}$, and has a variance of order $\frac{(2\Delta -1)^{1/2}}{\eta(t)}$. Note that the mean and the variance are of the same order, at least for small $\Delta$, but as $\Delta$ becomes large, the state becomes supported far from the boundary at $n=0$. Thus, we expect our approximation to become better at large $\Delta$.

Motivated by these observations, we define the rescaled Krylov index:
\beq 
x = \eta(t)\,n,
\eeq 
which behaves like a continuous variable in the $\eta \to 0$ limit. In this variable, the wavefunction can be approximated as
\beq 
\psi(x,t) = \frac{1}{\sqrt{\eta}}\psi_n(t) = \left(\frac{g^2}{\eta}\right)^{\Delta}\frac{1}{\sqrt{\Gamma(2\Delta)}} x^{\Delta -\frac{1}{2}}e^{-x} e^{i\frac{x}{\eta}\phi}.
\eeq 
Since the wavefunction spreads out in the Krylov index, it gets squeezed in the momentum direction $\theta$.  This also has the consequence that the Wigner function will be highly peaked in the momentum direction $\theta$.  

Note from equation \eqref{WigKry2} that under shifts $\theta \to \theta + \pi$, the Wigner function for integer $q$  is invariant  while for half-integer $q$ it  picks up an overall minus sign. 
Below we will show that the Wigner function  for integer $q$  develops two peaks at $\theta = -\phi$ and $\theta = -\phi + \pi$, while for half-integer $q$ it develops a peak at $\theta = -\phi$ and an anti-peak at $\theta = -\phi + \pi$.  Finally, we note that in all the examples considered above,
it is easy to check from equation \eqref{eq:fandg} that the phase of $f$, $\phi(t) \to \pi$ at late times.

We first focus on the Wigner function near $\theta = 0$ and for 
 $q$ being a non-negative integer.  As will become clear below, all the peaks and anti-peaks of the Wigner function contribute similarly to the Wigner negativity, and so including them all results in an overall factor of 4. Now note from equation \eqref{WigKry2} that in the momentum direction, the Wigner function $W(q,\theta)$ at most only involves Fourier modes of frequency up to $2q$. Furthermore, the wavefunction in position is peaked at $q\sim \frac{2\Delta-1}{\eta}$, and so the Wigner function is varying in the $\theta$ direction at scales of the order of $\eta$ which sets the width of the peak/anti-peak, but the function does not vary much at smaller momentum scales. Thus, we are led to define a  rescaled momentum variable around $\theta = 0$: 
\beq 
p = \frac{1}{\eta(t)}\,\theta,
\eeq 
where the rescaled momentum  lies in the range $p\in (-\frac{\pi}{2\eta}, \frac{\pi}{2\eta})$, and in the $\eta \to 0$ limit becomes an effectively non-compact variable. 

In the $\eta \to 0$ limit, we can approximate the Wigner function as
\beq 
W_{\text{eff}}(x,p) = W(q=x/\eta(t),\theta=\eta(t) p ),
\eeq 
where 
\beq 
W_{\text{eff}}(x,p) \simeq \frac{1}{\Gamma(2\Delta)}|g|^{4\Delta}\left(\frac{2x}{\eta}\right)^{2\Delta}e^{-2x}\int_0^{1}\frac{dy}{2\pi}\,e^{2i(2y-1)xp}\,\left(y(1-y)\right)^{\Delta-1/2}.
\eeq 
The $y$ integral can be done and we get
\beq 
W_{\text{eff}}(x,p) = \frac{1}{2}\left(\frac{|g|^2}{2\eta}\right)^{2\Delta}\left[\frac{1}{\Gamma(2\Delta)}(2x)^{2\Delta-1}e^{-2x}\right] \times \left[ \frac{2^{\Delta-1} \Gamma(\Delta + \frac{1}{2})}{\sqrt{\pi}}\,(2x)\,\frac{J_{\Delta}(2xp)}{(2xp)^\Delta}\right].
\eeq 
In the large $t$ limit $\frac{|g|^2}{2\eta}\to 1$, and so all the $t$-dependence  disappears from the Wigner function and has been absorbed into the limits on the momentum coordinate which lies in the range  $p\in (-\frac{\pi}{2\eta}, \frac{\pi}{2\eta})$. 

We can write the above formula for the effective Wigner function in the vicinity of the $\theta=0$  in a suggestive manner:
\beq \label{prod}
W_{\text{eff}}(x,p) = \frac{1}{2}w(x)w(p|x),
\eeq 
where 
\beq 
w(x) = \frac{2^{2\Delta}}{\Gamma(2\Delta)}x^{2\Delta-1}e^{-2x},\;\;
\eeq 
is the probability density in $x$ and
\beq 
w(p|x) = \frac{2^{\Delta-1}\Gamma(\Delta + \frac{1}{2})}{\sqrt{\pi}}\,(2x)\, \frac{J_{\Delta}(2xp)}{(2xp)^{\Delta}},
\eeq 
can be interpreted as a ``conditional quasi-probability'' distribution for $p$ given $x$.  The Bessel function above indicates that $W_{\text{eff}}(x,p)$ is peaked at $p=0$ and hence at $\theta =0$. The factor of $\frac{1}{2}$ in equation \eqref{prod} comes about because we are focusing on the vicinity of one of the two peaks in the $\theta$-direction. While $w(p|x)$ is approximately normalized, and precisely so in the $\eta \to 0$ limit,
\beq 
\lim_{\eta \to 0} \int_{-\frac{\pi}{2\eta}}^{\frac{\pi}{2\eta}} dp\,w(p|x) = 1,
\eeq 
notice that it is not positive everywhere. 

We can now calculate the Wigner negativity from the effective Wigner function as follows. Firstly, let's consider how the normalization works:
\beq 
\sum_{q=0,\frac{1}{2},1,\cdots} \int d\theta \,W(q,\theta) = 1.
\eeq 
For $q \in \mathbb{Z}$, we get a factor of $\frac{1}{2}$ from each of the two peaks at $\theta = 0$ and $\theta = \pi$ which add up to give one. For $q  \in \mathbb{Z} + 1/2$ we get a $\frac{1}{2}$ from the peak at $\theta = \pi$, but a $-\frac{1}{2}$ from the peak at $\theta = 0$, and so the half-integer $q$ points do not contribute to the total probability. 

Now let us consider the negativity. In this case, we get a $\frac{1}{2}$ from each peak both when $q$ is an integer and when it is a half-integer. So we find
\beqn 
\mathcal{N}(t) &\simeq & 2 \int_0^{\infty}dx\,w(x) \int_{-\frac{\pi}{2\eta}}^{\frac{\pi}{2\eta}} dp\,|w(p|x)|.
\eeqn 
Now consider the integral: 
\beq 
\mathcal{I}_\Delta(x) = \int_{-\frac{\pi}{2\eta}}^{\frac{\pi}{2\eta}} dp\,|w(p|x)|.
\eeq 
\begin{figure}[t]
 \centering
 \includegraphics[width=0.6\textwidth]{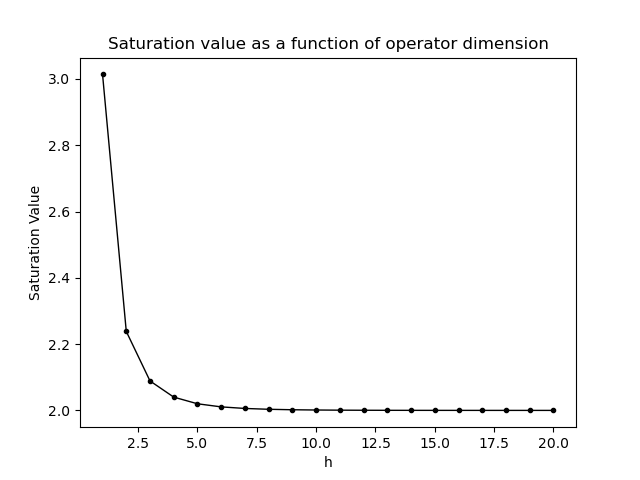}
    \caption{Numerical evaluation of the saturation value of the integral $\mathcal{I}_{\Delta}$ in the $\eta \to 0$ limit as a function of $\Delta =h$.}  
    \label{fig:sat_value}
\end{figure}
The integral $\mathcal{I}_\Delta$ converges in the $\eta\to 0$ limit for any $\Delta > 1/2$, and so in this case we find that the negativity saturates to a constant at late times in all the cases 1,2 and 3 (see Fig. \ref{fig:sat_value}):
\beq \label{WNsat}
\lim_{t\to \infty} \mathcal{N} = 2\int_{-\infty}^{\infty} dp\,|w(p|x)|.
\eeq 
This result agrees with what we saw in the previous sub-section by direct numerical computation (see figure \ref{fig:case_1}, \ref{fig:case_2} and \ref{fig:case_3}). 

Note that when $\Delta=\frac{1}{2}$, the integral $\mathcal{I}_{1/2}$ grows logarithmically in the $\eta \to 0$ limit. In this case,  negativity grows as $\log \frac{1}{\eta(t)}$ at late times (still much slower than the linear in $\eta$ growth of spread complexity). For $\Delta < \frac{1}{2}$, the growth is $\frac{1}{\eta^{\frac{1}{2}-\Delta}}.$ Thus, for $\Delta \leq \frac{1}{2}$, we see signs of non-classicality in the dynamics. On the other hand, as $\Delta \to \infty$, the late-time saturation value of the negativity approaches the value $2$. This is easy to understand -- as $\Delta$ becomes large, the function $w(p|x)$ approaches a Gaussian and has no negativity.  Thus, the state becomes effectively classical in this limit. This fits well with intuition from AdS/CFT, where in the large $\Delta$ limit, the bulk excitation can be well-approximated as a classical particle moving along a timelike geodesic
\cite{holographicParticle,Nozaki:2013wia}. 

We conclude with a remark on the probability distribution in momentum space. We saw previously (Fig.~\ref{fig:mom SL(2,R) h=2}) that the momentum space probability distribution $|\widetilde{\psi}(\theta,t)|^2$ develops a peak only around $\theta=\pi$. On the other hand, the Wigner function seems to develop a peak at $\theta=\pi$ and a peak/anti-peak at $\theta=0$, depending on whether $q$ is integer or half-integer. At first sight this might seem puzzling, but this  can be reconciled because the momentum space probability is given by:
\beq
|\widetilde{\psi}(\theta,t)|^2=\sum_{q=0,\frac12,1...} W(q,\theta).
\eeq
Since the sum is over both integer and half-integer values of $q$, the peaks at $\theta=0$ for integer $q$ get canceled by the anti-peaks from half-integer $q$. On the other hand, at $\theta=\pi$ the contributions from integer and half-integer $q$ add up, giving rise to a peak in $|\widetilde{\psi}|^2$.

\subsection{Summary}
In summary, we have shown that for operator dimension $\Delta  > \frac{1}{2}$, while the wavefunction in the Krylov basis spreads out rapidly under time evolution as evident in the spread complexity, the negativity of the Wigner function does not grow, and in fact saturates to a constant. This means that despite the spread of the wavefunction, the state and dynamics remain approximately classical.

\section{Random matrix theory} \label{sec:RMT}
\subsection{Krylov effective description}
Let the Hamiltonian be  a $D\times D$ random matrix drawn from the Gaussian unitary ensemble (GUE) with the probability measure:
\beq 
\mu(H) =\frac{1}{Z} e^{-\frac{D}{2\epsilon^2}\,\mathrm{Tr}\,H^2}.
\eeq 
Henceforth, we  set $\epsilon=1$ and measure all dimensionful quantities (such as energy, time etc.) in these units. Take the initial state to be the maximally-entangled state $|\Omega\rangle$ between two copies of the system. We wish to study the time evolved state:
\beq 
|\psi(t)\rangle = e^{-itH}|\Omega\rangle,
\eeq 
in the Krylov basis. Remarkably, for Hamiltonians drawn randomly from the GUE,  the Krylov coefficients $\{a_n\}$ and $\{b_n\}$ behave as independent random variables with normal and $\chi$ distributions respectively \cite{10.1063/1.1507823, Balasubramanian:2022dnj}. In the large-$D$ limit, their average values are given by
\beq \label{meanKCs}
\overline{a_n} = 0, \qquad\overline{b_n} = 1\,,\qquad (n\;\text{fixed},D\to \infty),
\eeq 
with standard deviations of $O(1/\sqrt{D})$. Thus, we get an \emph{effective Hamiltonian} for dynamics in the subspace with $n$ held fixed as $D\to \infty$: 
\beq \label{Heff}
H_{\text{eff}}|n\rangle = |n+1\rangle + |n-1\rangle.
\eeq 
This effective Hamiltonian is sufficient to describe time evolution at sub-exponential times \cite{Basu:2025mmm, Hochbruck}.

The eigenstates of this effective Hamiltonian are well-known \cite{SusskindGlogower, phase} (see also Appendix E of \cite{Rabinovici:2023yex}):
\beq 
H_{\text{eff}} |\theta\rangle =2\cos\,\theta |\theta\rangle\,,\qquad|\theta\rangle = \sqrt{\frac{2}{\pi}}\sum_{n=0}^{\infty}\sin[(n+1)\theta]|n\rangle\,.
\eeq
They satisfy the completeness relation:
\beq 
\int_0^{\pi}d\theta\,|\theta\rangle \langle \theta| = 1.
\eeq 
We can thus compute the transition amplitude $\langle k | e^{-itH}|0\rangle$ in the large  $D$ limit and we find:
\beq\label{eq:aly}
\psi_k(t)= \langle k|e^{-itH_{\text{eff}}}|0\rangle = \frac{1}{i^k}\frac{(k+1)}{t}J_{k+1}(2t).
\eeq
Using properties of the Bessel function, we see that the wavefunction is oscillatory for $k\leq 2t$ and decays exponentially for $k>2t$.  The  spread complexity of these wave functions can be found e.g. in \cite{Barbon:2019wsy,Rabinovici:2023yex}, and grows linearly at late times.

\subsection{Wigner negativity}
\begin{figure}[t]
 \centering
 \begin{tabular}{c c}
 \includegraphics[width=0.45\textwidth]{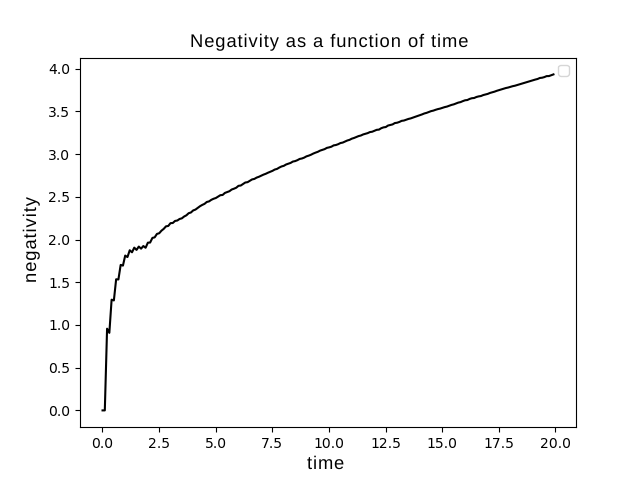} & \includegraphics[width=0.45\textwidth]{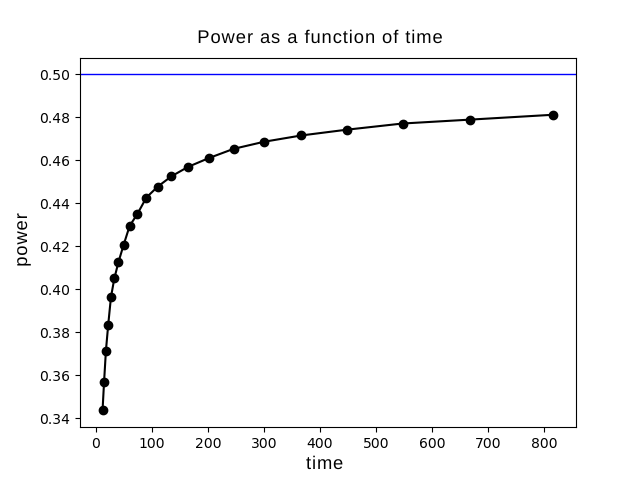}
 \end{tabular}
    \caption{\textbf{(Left)} Negativity as a function of time in the large-$D$ limit for random matrix theory with Gaussian unitary ensemble. \textbf{(Right)} The power $t\frac{d}{dt}\log\mathcal{N}$, where $\mathcal{N}$ is the negativity as a function of time.   }  
    \label{fig:neg_gue}
\end{figure}

We can now obtain the Wigner function as:
\beq 
W(q,\theta) = \frac{1}{2\pi}\sum_{m,n=0}^{\infty}e^{i\theta(m-n)}\delta_{2q,m+n}\psi_m(t)\psi^*_n(t),
\eeq
and compute its negativity as a function of time. It is challenging to get analytic expressions, but in Fig.~\ref{fig:neg_gue}, we plot the Wigner negativity numerically and find evidence for power-law growth $t^{\gamma}$ at late times, with the power being $\gamma \simeq \frac{1}{2}$.  In \cite{Basu:2025mmm}, it was shown that the negativity growth is bounded above by $\sqrt{t}$; our numerical results suggest that this bound is in fact saturated at late times. 

The $\sqrt{t}$ growth at late times can be understood heuristically by evaluating the Wigner function in the saddle point approximation. To this end, consider the limits $D\to \infty$ followed by $t\to \infty,$ in that order. From the integral representation of the Bessel function:
\beq 
J_{k+1}(2t) = \frac{i^{-k}}{\pi}\int_0^\pi d\theta \,e^{2it\cos\theta}\cos[(k+1)\theta],
\eeq 
we see that the $t \to \infty $ limit looks like an ``$\hbar \to 0$'' limit, where the exponential becomes  oscillatory.  In this limit  we can apply the stationary phase approximation. Of course, $k$ runs over non-negative integers, and can itself scale with $t$. Therefore, a natural limit to consider is $t\to \infty,\,k\to \infty$ with $x= \frac{k+1}{2t}$ fixed to be an arbitrary, non-negative real number. We can think of this as a ``continuum'' approximation of the wavefunction $\langle k|e^{-itH_{\text{eff}}}|0\rangle$, where instead of the discrete Krylov lattice, we now have a wavefunction of a continuous parameter $x$. In this limit, it is well-known that the Bessel function $J_{\nu}(z)$ admits a Debye expansion, where $\nu \to \infty, z\to \infty$ with $\nu/z$ held fixed.   In the range $0 < \frac{k}{2t}<1$, this expansion gives the oscillatory behavior:
\beq
J_{k+1}(2t) \sim  \left(\frac{2}{2\pi t \sin(\beta)}\right)^{1/2}\cos\left[2t(\sin\,\beta - \beta \cos\,\beta)- \frac{\pi}{4}\right]\;\;\cdots \;\;(\cos\,\beta =\frac{k+1}{2t}),
\eeq
while in the range $\frac{k}{2t}>1$, the analogous formula shows that the Bessel function decays exponentially in $k$:
\beq
J_{k+1}(2t) \sim \frac{1}{\sqrt{4 \pi t \sinh \alpha}} \exp\Big[-2t(\alpha \cosh \alpha- \sinh \alpha )\Big]\;\; \cdots\;\; (\cosh\alpha= \frac{k+1}{2t}).
\eeq
In terms of the rescaled variable $x$, the wavefunction is thus given by:
\beq
\psi(x,t)=\begin{cases} 2ix e^{-i \pi x t} \left(\frac{1}{\pi t \sqrt{1-x^2}}\right)^{1/2}\text{cos}\left[2 t \left(\sqrt{1-x^2}-x\, \cos^{-1}(x)\right)-\frac{\pi}{4}\right] \;\;\; \cdots \;(x<1)\,, \\ 
 2ix e^{-i \pi x t} \left(\frac{1}{4 \pi t \sqrt{x^2-1}}\right)^{1/2} \exp\left[-2t\left(x \cosh^{-1}(x)-\sqrt{x^2-1}\right] \right] \;\;\; \cdots \;(x>1)\,.\end{cases}
\eeq
These formulas show that the wavefunction is well-localized within the region $k \leq 2t$ where it is oscillatory, and decays exponentially beyond $k=2t$. The wavefunction in the transition region close to $x=1$ can be expressed in terms of the Airy function, but we will not need to use this here.

Coming back to the Wigner function, we have 
\beq\label{eq:WigBess}
W(q,\theta)=\frac{1}{2 \pi} \sum_{m=0}^{2q} e^{2i(m-q)(\theta-\frac{\pi}{2})} \frac{(m+1)(2q-m+1)}{t^2} J_{m+1}(2t) J_{2q-m+1}(2t).
\eeq
As in the previous section, shifting $\theta \to \theta + \pi$ leaves the Wigner function invariant for $q$ being an integer, and gives an overall sign in the Wigner function for   half-integer $q$. For this reason, it is sufficient to study the Wigner function in a fundamental domain in momentum space that we will take as $\theta \in (0, \pi)$, and separately consider the cases where $q$ is either integer or half integer valued.  We will for now restrict attention to the case where $q$ is integer valued, and $\theta \in (0, \pi)$ with the understanding that we get an overall factor of $4$ in the negativity from accounting for the remaining regions in phase space.  Within this sector, we define the following rescaled position variables:
\beq
y = \frac{m+1}{2t}\,,\qquad x = \frac{q+1}{2t}\,.
\eeq 
In the large-$t$ limit, the sum over $m$ can be approximated by an integral over $y$. The Wigner function is primarily localized in the region $ 0 \leq x \leq 1$, because outside this region one of the Bessel functions in equation \eqref{eq:WigBess} decays exponentially. Furthermore, if we keep only the regions in $y$ where the wavefunctions are both oscillatory (as opposed to exponentially decaying) we must  have
\beq 
0 \leq y \leq 1\,, \qquad 0 \leq 2x-y \leq 1\,,
\eeq 
which implies 
\beq 
\text{max}(0,2x-1) \leq y \leq \text{min}(1, 2x)\,.
\eeq 
Putting everything together, the Wigner function is given by:
\beq \label{eq:WigFunc}
W(x,\phi) = \sum_{\sigma,\sigma'} \int_{\text{max}(0, 2x-1)}^{\text{min}(1,2x)} dy \  g_{\sigma,\sigma'}(y) e^{2i t f_{\sigma,\sigma'}(y)}\,,
\eeq
where $\sigma,\sigma' = \pm 1$, and
\begin{eqnarray*}
g_{\sigma,\sigma'}(y)&=& \frac{1}{\pi^2}\frac{y(2x-y)}{(1-y^2)^{\frac{1}{4}}(1-(2x-y)^2)^{\frac{1}{4}}} e^{-i\frac{\pi}{4}(\sigma + \sigma')},\\
f_{\sigma,\sigma'}(y)&=& 2\phi (y-x)+ \sigma \left(\sqrt{1-y^2}-y\,\text{cos}^{-1}y\right) \nonumber\\
&+& \sigma' \left(\sqrt{1-(2x-y)^2}-(2x-y)\text{cos}^{-1}(2x-y)\right),
\end{eqnarray*}
with $\phi = (\theta-\frac{\pi}{2})$. 

We can  perform these integrals in the saddle point approximation. The saddle point equations are:
\beq\label{eq:stationary_phase}
f_{\sigma,\sigma'}'(y)=0 \;\;\Rightarrow \;\;
\sigma\, \text{cos}^{-1}(y^*) - \sigma' \text{cos}^{-1}(2x-y^*)-2\phi=0.
\eeq
These expressions reveal why the negativity should grow as $t^{1/2}$. First, we know that the saddle point contributions to the Wigner function will be of $O(t^{-1/2})$, where the $t^{-1/2}$ comes from the one-loop determinant around each saddle determined by equation \eqref{eq:stationary_phase}.  Furthermore, the formula for the negativity has a sum over $q$ which gives a factor of $t$ coming from the fact that the wavefunction is oscillatory in the range $0 \leq n \leq 2t$.   Finally, the negativity density (i.e., the amount of negativity in a small $\epsilon$-window in the $q$ direction) is $O(1)$. This is because the only other $t$-dependence lies in the factor $e^{2it f_{\sigma,\sigma'}(y^*)}$ in equation \eqref{eq:WigFunc}, and a function that oscillates with a frequency of $t$ has a negativity density of $O(1)$.\footnote{Consider for instance, the oscillatory function $\cos(tx)$. Then $\int_{x_0- \epsilon}^{x_0 + \epsilon}dx |\cos(t\,x)|$ is proportional to the number of periods within the range $(x_0-\epsilon,x_0+\epsilon)$ -- which is proportional to $t$, times the integral within one period which is $\frac{1}{t}\int_0^{2\pi}d\theta \,|\cos(\theta)|$ (where we have defined $x = \theta/t)$ -- which is $O(\frac{1}{t}).$}  Thus, we expect that the negativity coming from the saddle point contributions should be of $O(t^{1/2})$.   

\begin{figure}[t]
 \centering
 \begin{tabular}{c c}
 \includegraphics[width=0.48\textwidth]{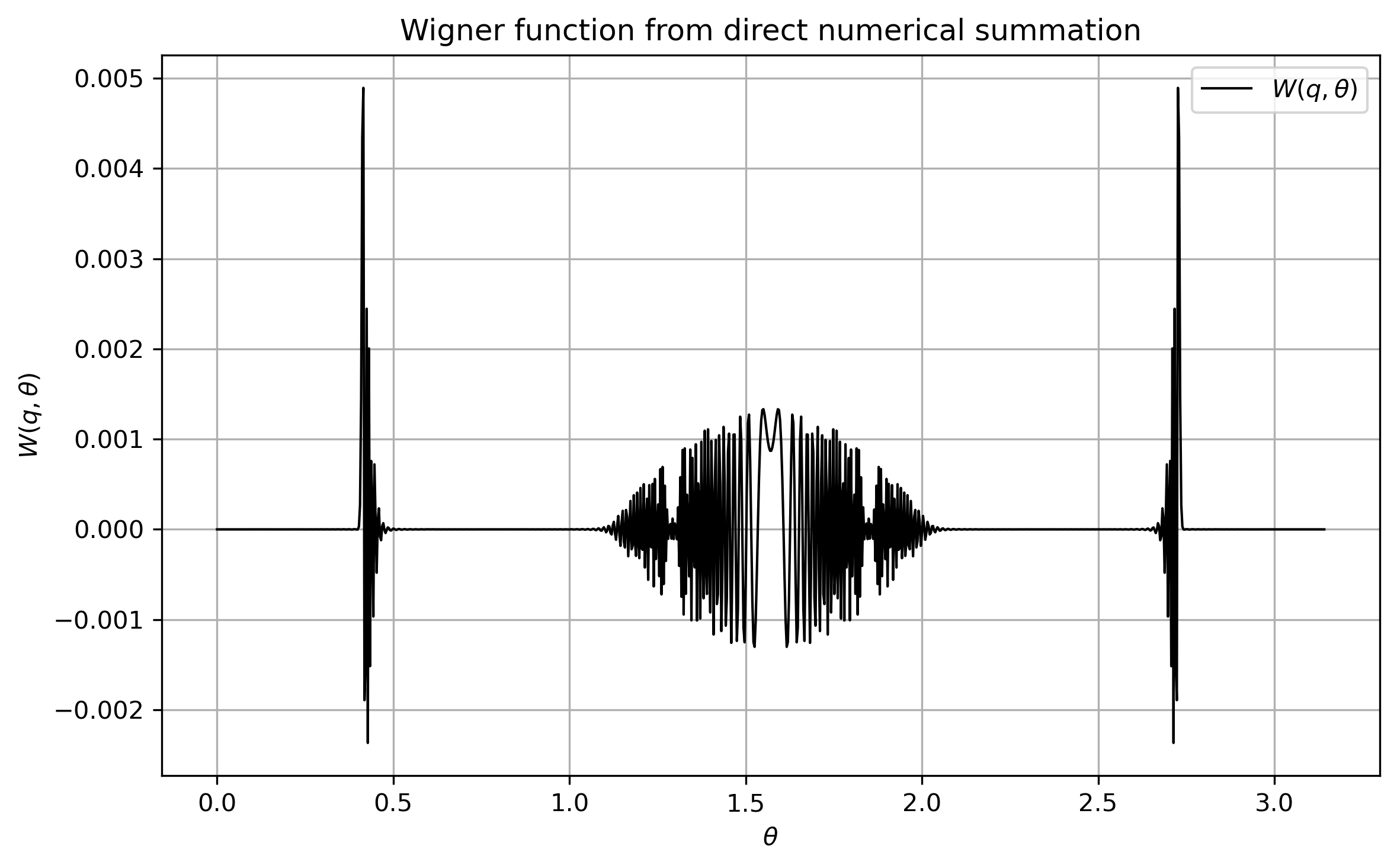} & \includegraphics[width=0.48\textwidth]{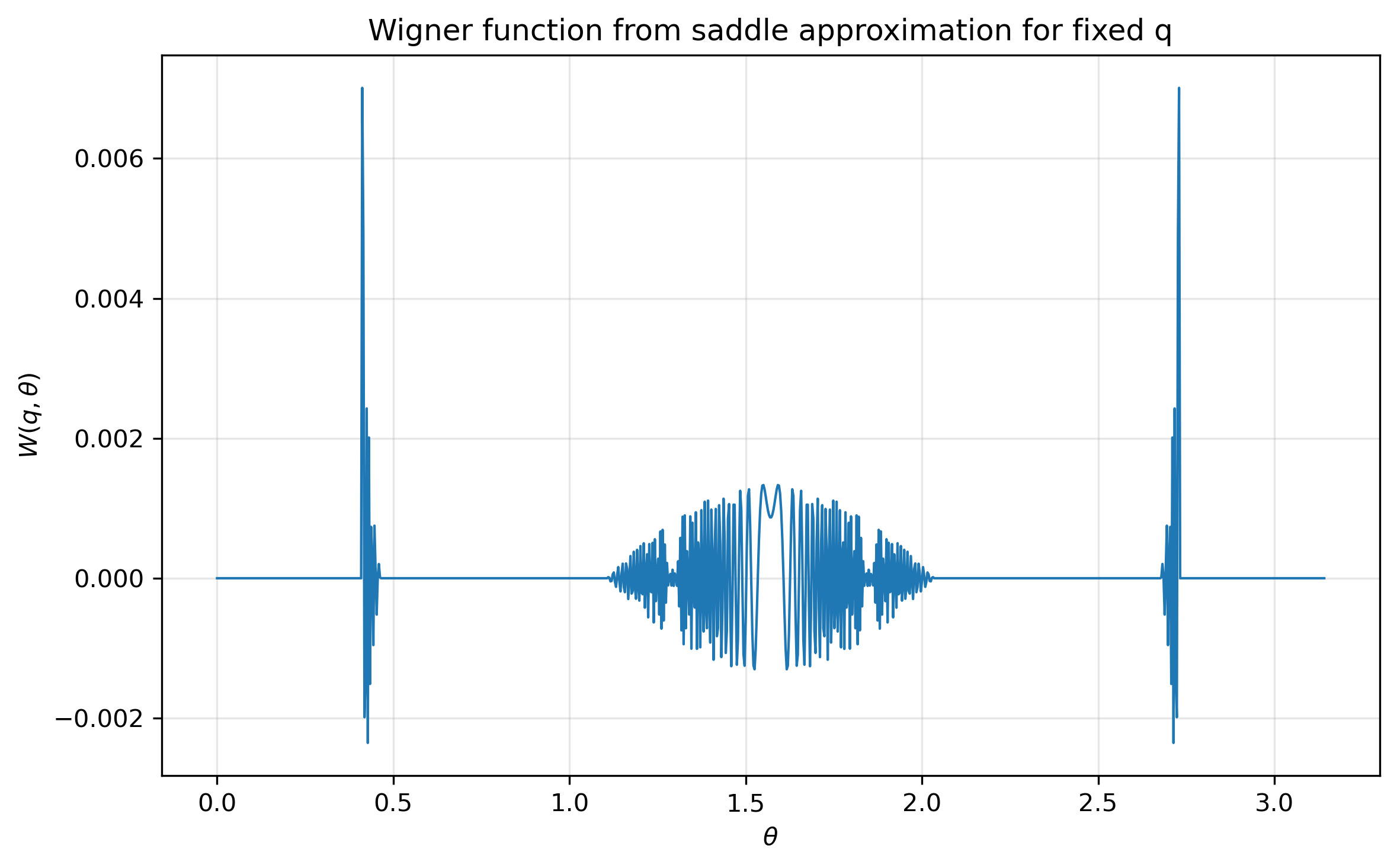}
 \end{tabular}
    \caption{Exact numerical evaluation (left) v. leading stationary-phase approximation (right) of the Wigner function $W(q,\theta)$ at $q=799$ and $t=1000$ as a function of $\theta$}  
    \label{fig:wig_gue_comparison}
\end{figure}

\begin{figure}[t]
 \centering
 \begin{tabular}{c c}
 \includegraphics[width=0.48\textwidth]{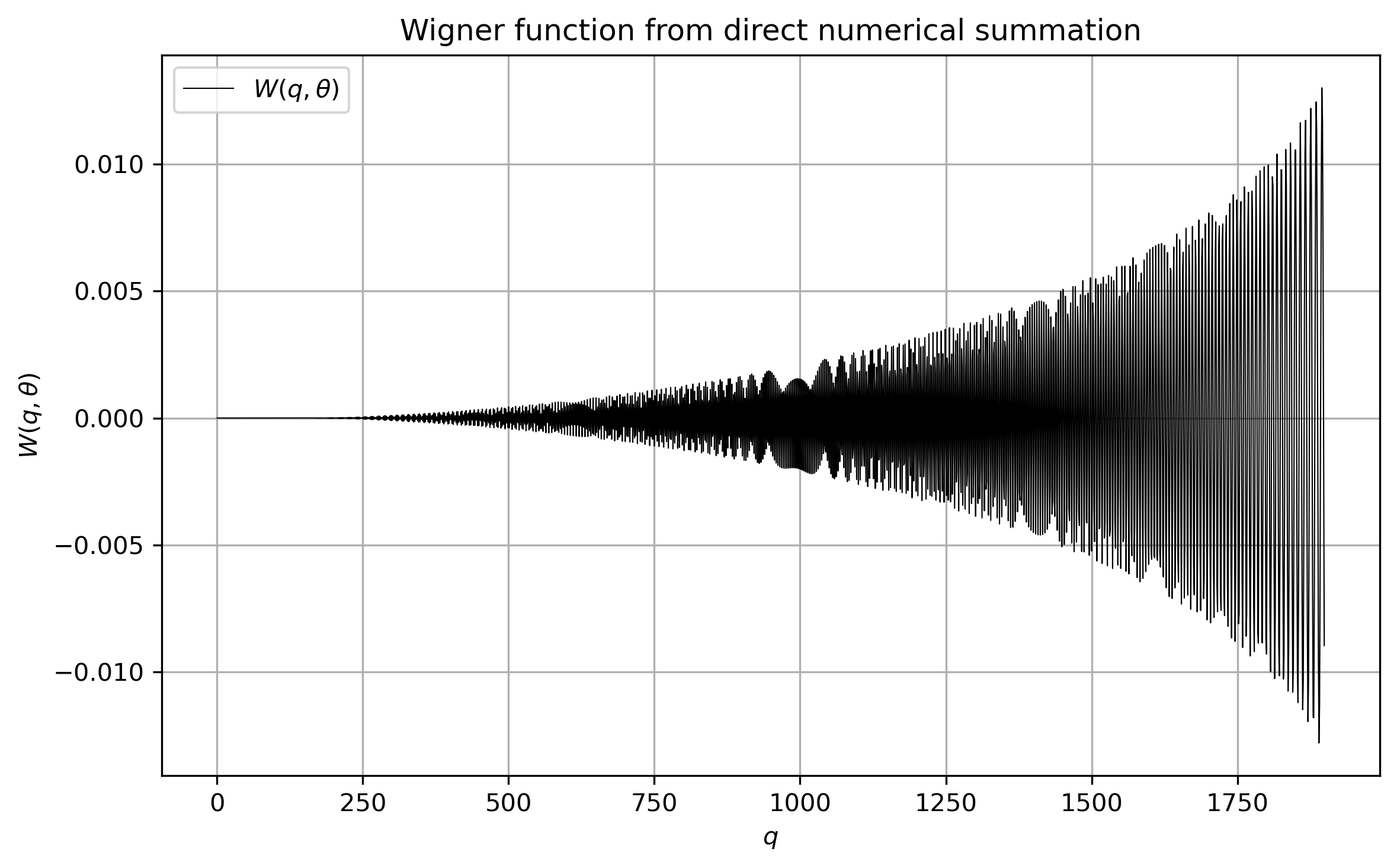} & \includegraphics[width=0.48\textwidth]{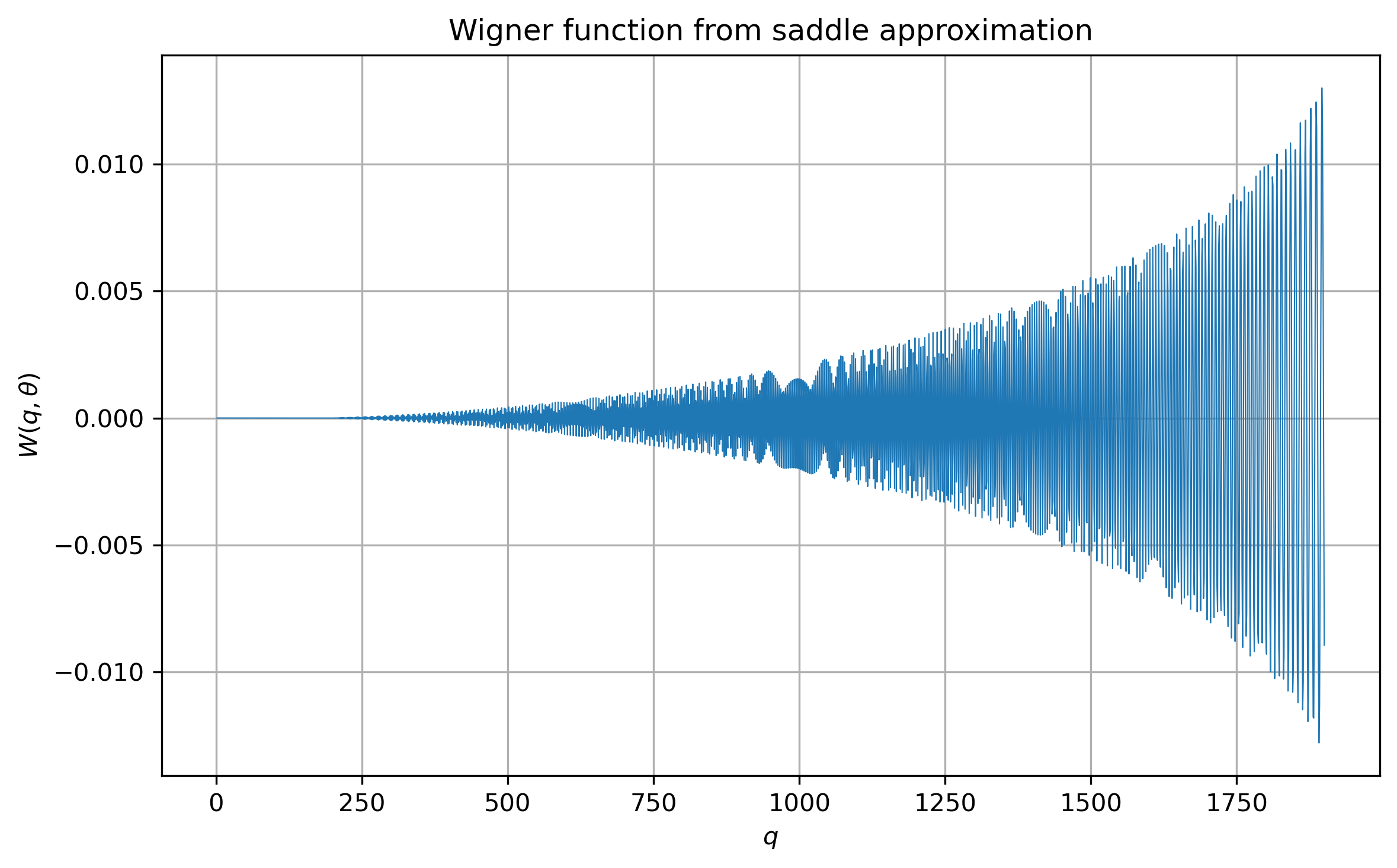} 
 \end{tabular}
    \caption{Exact numerical evaluation (left) v. leading stationary-phase approximation (right) of the Wigner function $W(q,\theta)$ at $\theta=\pi/2+0.1$ and $t=1000$ as a function of $q$.}  
    \label{fig:wig_gue_comparison_theta_fixed}
\end{figure}

\begin{figure}[t]
 \centering
 \begin{tabular}{c c}
 \includegraphics[width=0.45\textwidth]{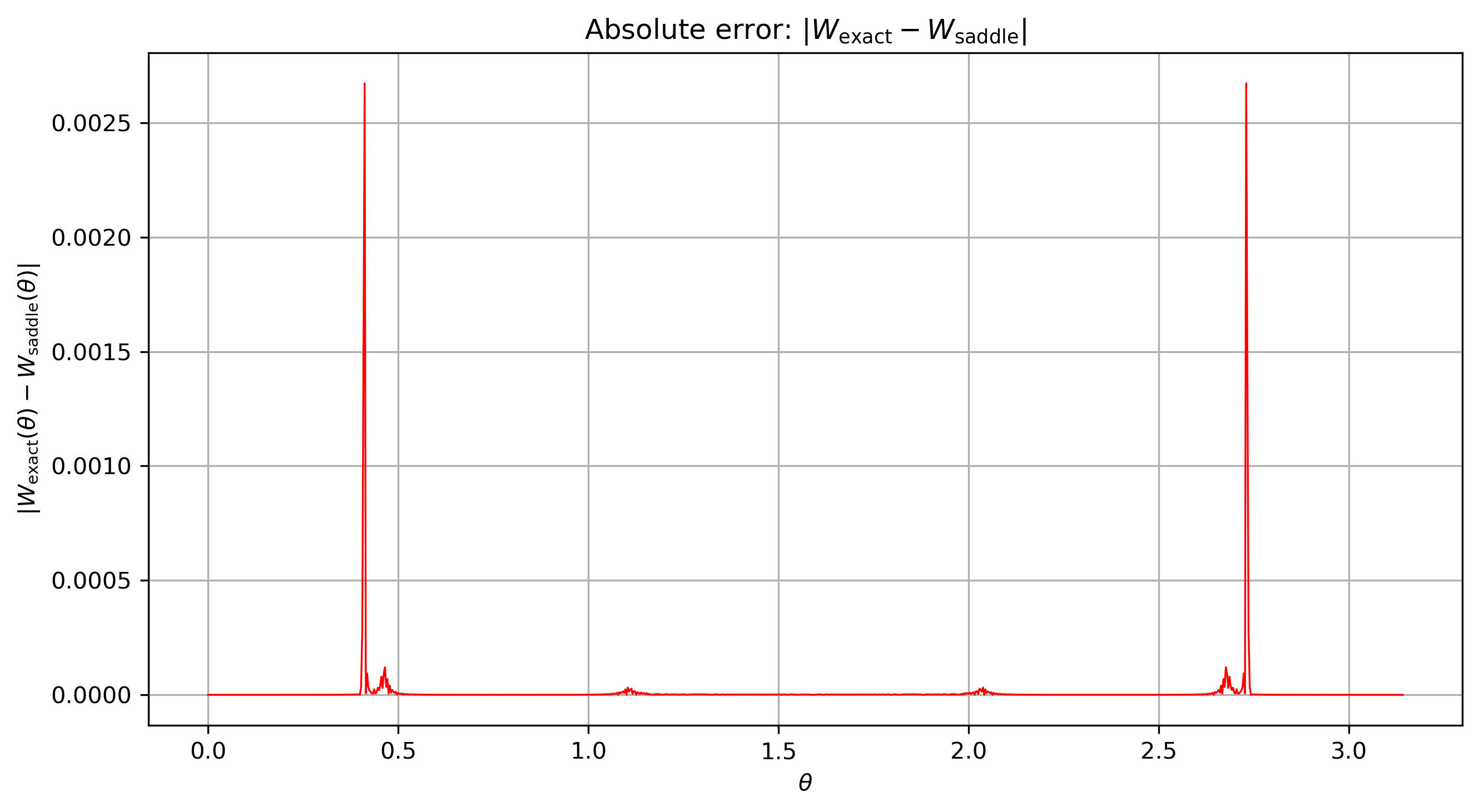} & \includegraphics[width=0.45\textwidth]{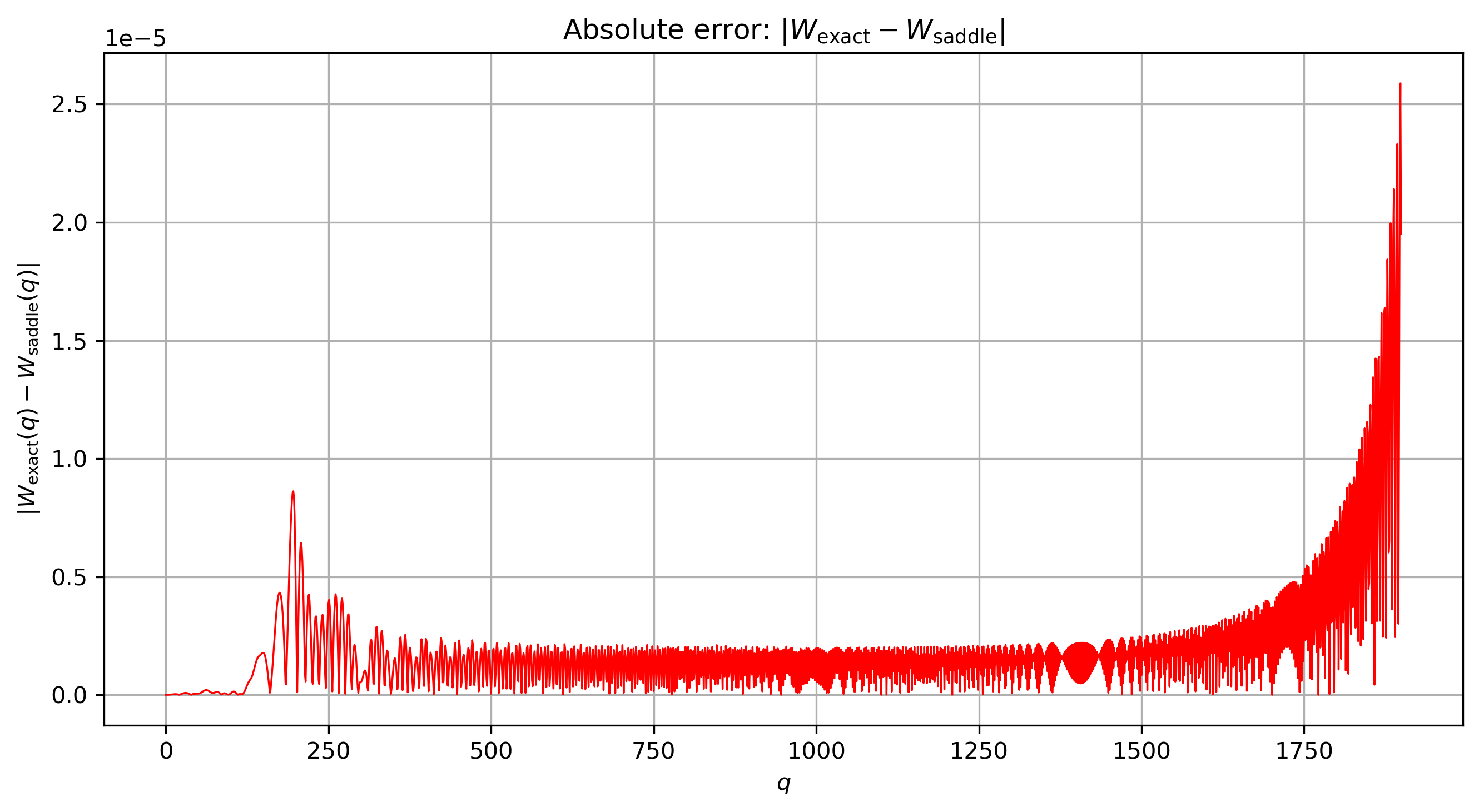} 
 \end{tabular}
    \caption{Absolute error $\left|W_{\mathrm{exact}}-W_{\mathrm{saddle}}\right|$ between the direct numerical evaluation of the Wigner function and the saddle-point approximation at $t=1000$: \textbf{(Left)} \ as a function of $\theta$ for a fixed value of $q=799$; \textbf{(Right)} as a function of $q$ for fixed $\theta=\pi/2+0.1$. The error is small except at some special points. These are the transition points between regions where the solution to saddle point equation exists vs the region where it does not exist. The error near these transition points is large since a more careful analysis is needed near these transition points but the corrections are sub-leading for the Wigner negativity calculation.}  
    \label{fig:wig_gue_error}
\end{figure}

Having argued why negativity should grow at $t^{1/2}$, let us calculate what the Wigner function would be in the large-$t$ limit.  In Table~\ref{tab:sol} we summarize solutions to the saddle point equation \eqref{eq:stationary_phase} and the range of validity for each of the values of $\sigma, \sigma'$.  (See details in App.~\ref{app:stationary}). Using these saddle point solutions we can explicitly write down the large-$t$ Wigner function as:
\beqn
W(x,\phi)=\sum_{\sigma,\sigma'} \sqrt{\frac{\pi}{t |f''_{\sigma,\sigma'}(y^*)|}} \ g_{\sigma,\sigma'}(y^*) \ e^{2 i t f_{\sigma,\sigma'}(y^*)+\frac{i\pi}{4} sgn(f''_{\sigma,\sigma'}(y^*))}\,.
\eeqn
Figs.~\ref{fig:wig_gue_comparison} and \ref{fig:wig_gue_comparison_theta_fixed} show this large-$t$ Wigner function obtained by the stationary phase method, compared with the exact Wigner function calculated numerically. In Fig.~\ref{fig:wig_gue_error}, we also plot the absolute error $\left|W_{\mathrm{exact}}-W_{\mathrm{saddle}}\right|$, for both fixed $q$ and fixed $\theta$,  and see that the two are in excellent agreement except for a few points where transitions  occur whereby saddle points  appear or disappear. 

\begin{table}[h!]
\centering
\small
\setlength{\tabcolsep}{4pt}
\begin{tabularx}{\columnwidth}{|c|X|c|c|}
\hline
$\sigma,\sigma'$&\centering solution &r.o.v for $x<1/2$& r.o.v. for $x>1/2$
\\
\hline
$1,1$&\centering$y^*=x-\tan\phi\,\sqrt{\cos^2\phi-x^2}$
&
$\displaystyle |\phi|<\pi/4-h_1(x)$
&
$\displaystyle |\phi|<h_2(x)$
\\
\hline
$-1,-1$
&
\centering$y^*=x+\tan\phi\,\sqrt{\cos^2\phi-x^2}$
&
$\displaystyle |\phi|<\pi/4-h_1(x)$
&
$\displaystyle |\phi|<h_2(x)$
\\
\hline
$1,-1$
&\centering$y^*=x\pm\tan\phi\,\sqrt{\cos^2\phi-x^2}$
&$\displaystyle \pi/4+h_1(x)<\phi<h_3(x)$
&
$\displaystyle h_2(x)<\phi<h_3(x)$
\\
\hline
$-1,1$
&\centering
$y^*=x\pm\tan\phi\,\sqrt{\cos^2\phi-x^2}$
&
$\displaystyle \pi/4+h_1(x)<-\phi<h_3(x)$
&
$\displaystyle h_2(x)<-\phi<h_3(x)$
\\
\hline
\end{tabularx}
\caption{Solutions and range of validity (r.o.v) for the stationary phase condition.  Here $h_1(x)=\frac{\text{cos}^{-1}(2x)}{2}$, $h_2(x)=\frac{\text{cos}^{-1}(2x-1)}{2}$ and $h_3(x)=\text{cos}^{-1}(x)$ }
\label{tab:sol}
\end{table}

To summarize, we have argued that at late but $O(1)$ times (i.e., not scaling with the dimension of the Hilbert space), the Wigner negativity in the Krylov basis grows as $t^{1/2}$. Thus, the Krylov basis provides a \emph{semiclassical} description of  time evolution at all $O(1)$ times, even though the Hamiltonian is a random matrix. This is reminiscent of the AdS/CFT correspondence, where the bulk gravitational theory provides a semiclassical description of the chaotic boundary dynamics. 
\subsection{Beyond GUE}

While the analysis in this section was performed for the  GUE, we expect that the $\sqrt{t}$ growth of the Wigner negativity at late times will be a  universal feature of all random matrix models where the large-$D$ density of states is compactly supported on an interval. This is because in all such matrix models, the finite extent of support in energy implies that the Lanczos coefficients $\{a_n, b_n\}$ saturate asymptotically at large $n$ \cite{Kar:2021nbm, math_paper}. Thus, at sufficiently late times, we expect that the state starts exploring this asymptotically saturated region in the Lanczos spectrum. At that point, the dynamics should essentially be controlled by the effective Hamiltonian \eqref{Heff} that also arises from the GUE, and  the Wigner negativity should follow the same $\sqrt{t}$ growth. We will see evidence towards this in the next subsection where we study negativity growth in the DSSYK model.   

Finally, note that the analysis in this section was limited to the $D \to \infty$ limit. For finite $D$, the conclusions above should hold at all times $t \ll D$. As $t$ approaches $D$, however, the negativity growth will truncate and saturate at an exponentially large value \cite{2024JHEP...05..264B, Basu:2025mmm}:
\beq 
\mathcal{N} \sim \sqrt{\frac{2D}{\pi}},
\eeq 
signaling a breakdown of the semiclassical effective description in the Krylov basis. The above saturation value corresponds to the negativity of a typical state in the Hilbert space \cite{White:2020hgn}, and thus we learn that at exponential times, the time-evolved state looks like a random superposition of basis vectors even with respect to the Krylov basis. 
\subsection{Summary}
In summary, we have argued that the negativity of the Wigner function in RMT grows universally as a power law at large but $O(1)$ times. This means at any $O(1)$ time negativity does not scale with the Hilbert space dimension.  In this sense, following the nomenclature introduced in Sec.~\ref{sec:semiclassical}, the state remains semiclassical in the Krylov basis.

\section{Double-Scaled SYK model} \label{sec:DSSYK}
Finally, we study the time evolution of Wigner negativity in the double-scaled SYK model (DSSYK) \cite{Cotler:2016fpe, Berkooz:2018jqr,Berkooz:2018qkz}. The DSSYK model is a theory of $N$ Majorana fermions with the Hamiltonian comprising  random interactions between $p$ fermions, in the double scaling limit where $N \to \infty, p \to \infty$, with $\frac{2p^2}{N} = \lambda$ held fixed. In this limit,  the theory can be solved in terms of {\it chord diagrams}, which arise from the following effective Hamiltonian for {\it chords states}:
\beq
H_{\text{eff}}=-\frac{1}{\sqrt{\lambda}}\left(\alpha \sqrt{\frac{1-\fq^{\hat{n}}}{1-\fq}} +  \sqrt{\frac{1-\fq^{\hat{n}}}{1-\fq}} \alpha^{\dagger} \right),
\label{eq:Hdssyk}
\eeq 
where $\fq=e^{-\lambda}$ and the operators $\alpha$, $\alpha^{\dagger}$ and $\hat{n}$ are chord annihilation, chord creation and the chord number operators respectively:
\beq 
\alpha |n\rangle = |n-1\rangle, \qquad\alpha^{\dagger} |n\rangle = |n+1\rangle,\qquad \hat{n} |n\rangle = n|n\rangle.
\eeq 
It was pointed out in \cite{Lin:2022rbf} that the chord number basis precisely corresponds to the Krylov basis with the initial state taken to be the maximally entangled state on two copies of the Hilbert space, with the chord number being equal to the Krylov index. Indeed, notice that the  Hamiltonian above takes a tri-diagonal form:
\beq 
H_{\text{eff}}|n\rangle = a_n|n\rangle + b_{n}|n-1\rangle + b_{n+1}|n+1\rangle,
\eeq 
with 
\beq \label{eq:DSSYKLanczos}
a_n = 0,\qquad b_n = \sqrt{\frac{1- \fq^n}{1-\fq}}.
\eeq 
Spread complexity in DSSYK has already been extensively studied in the past (see the review \cite{Rabinovici:2025otw}) and the link with the wormhole length in JT gravity was established in \cite{Lin:2022rbf,Rabinovici:2023yex}.

The above features make the DSSYK model a natural toy model for the purposes of studying the time evolution of Wigner negativity in the Krylov basis. One important difference here from the case of 2d-CFTs is that the support of the density of states is confined to a compact range of energies unlike the CFT case, and so DSSYK is in this sense a UV complete model. As we will see, the dynamics of Wigner negativity in this model resembles that of 2d CFT at small times and has a similar classical phase, but departs at long times where the finite extent of the energy spectrum becomes important, and the dynamics becomes more RMT-like. 

We start with the initial state $\ket{0}$ corresponding to the maximally entangled state \cite{Berkooz:2018qkz}. In order to evaluate the time evolution of Wigner negativity we need to compute the wavefunction of the time-evolved state:
\beqn
\psi_n(t)=\bra{n}e^{iHt}\ket{0},
\eeqn
in the Krylov basis. This can be calculated using chord diagrams \cite{Lin:2022rbf,Rabinovici:2023yex} and is given by:
\beqn
\psi_n(t)=\int_0^{\pi} \frac{d\theta}{2 \pi}\rho(\theta) \frac{H_l(\text{cos}(\theta)|\fq)}{\sqrt{(\fq;\fq)_l}} e^{-itE(\theta)},
\eeqn
where
\beq 
\rho(\theta)=(\fq;\fq)_{\infty}(e^{2i\theta};\fq)_{\infty}(e^{-2i\theta};\fq)_{\infty},
\eeq
is the density of states with $E(\theta)=\frac{-2\text{cos}(\theta)}{\lambda \sqrt{1-\fq}}$, and
\beq 
(a;\fq)_{\infty} = \prod_{k=0}^{\infty}(1-a \fq^k).
\eeq 
The integral can be performed to write the wavefunction in terms of an infinite sum, which is useful for numerical calculations:
\beq
\psi_n(t)=\sum_{r=0}^{\infty} (-1)^r \fq^{\frac{1}{2}r(r+1)} \frac{(\fq;\fq)_{r+n}}{\sqrt{(\fq;\fq)_n} (\fq;\fq)_r} (2r+n+1) \frac{\sqrt{\lambda (1-\fq)}}{-i t} I_{2r+n+1}\left( \frac{-2it}{\sqrt{\lambda (1-\fq)}}\right).\label{eq:syk0par}
\eeq 
Before the numerical analysis, we will first analyze the Wigner negativity growth in two analytically tractable limits. Recall from equation \eqref{eq:DSSYKLanczos} that the Lanczos coefficients $b_n = \sqrt{\frac{1-\fq^n}{1-\fq}}$ grow as $b_n\sim \sqrt{n}$ for $n\lambda \ll 1$, while for $n\lambda \gg 1$, we get $b_n \sim \frac{1}{\sqrt{1-\fq}}$, which is essentially identical (up to an overall scaling) to the GUE Lanczos spectrum. If we send $\lambda \to 0$ while holding $n$ and $t$ fixed, then we land in the former regime, and we will think of this as early times. On the other hand, if we send $\lambda \to \infty$ while holding $n$ and $t$ fixed, then we land in the GUE regime, which we can think of as late (but $O(1)$) times.
\subsubsection*{$\lambda \to 0$: Heisenberg model}
If we define the rescaled Hamiltonian $H_{\text{new}} = \sqrt{\lambda} H_{\text{old}}$ and then take the $\lambda \to 0$ limit, then the rescaled Hamiltonian simplifies to: 
\beq
H=-\left( \alpha \sqrt{\hat{n}}+\sqrt{\hat{n}} \alpha^{\dagger}\right).
\eeq
The time evolved wavefunction in this limit is:
\beq
\psi_n(t)=\frac{e^{-\frac{t^2}{2}}}{\sqrt{n!}} (it)^n,
\eeq
which matches the exact solution for a particle moving on the  the Heisenberg-Weyl group manifold  \cite{Caputa:2021sib} (see also Appendix~\ref{App:HW}). 

At late times, the wavefunction is peaked at large values of $n$. In this case, using Sterling's approximation, the wavefunction can be approximately written as:
\beq 
\psi_n(t) \simeq \frac{1}{(2\pi n)^{1/4}}e^{-\frac{t^2}{2}}e^{i\frac{\pi}{2}n} e^{n \log\,t-\frac{1}{2}n\log\,n +\frac{n}{2}}.
\eeq
 We see that the probability density corresponding to this wavefunction is peaked at $n_* = t^2$, with a width proportional to $\delta n \sim n^{1/2}_*$. Thus, we are led to define the rescaled variable
\beq 
n = t^2 + t\,x,
\eeq 
in terms of which the wavefunction becomes
\beq 
\psi_n(t) \simeq \frac{1}{(2\pi t^2)^{1/4}}e^{i\frac{\pi}{2}(t^2 + tx)} e^{-\frac{1}{4}x^2}.
\eeq 

Coming to the Wigner function, recall from our definition that:
\beq 
W(q,\theta) = \frac{e^{-t^2}}{2\pi}\sum_{m=0}^{2q}e^{i(\theta+\frac{\pi}{2})(2m-2q)}\frac{t^{2q}}{\sqrt{m!(2q-m)!}}.\label{eq:WforH}
\eeq
As in previous sections, the phase above is invariant under $\theta \to \theta + \pi$ when $q$ is an integer, while it remains invariant up to an overall minus sign when $q$ is a half-integer. As a result, this Wigner function has two peaks in $\theta$ (separated by $\pi$) for integer $q$, while it should have a peak and an anti-peak when $q$ is half-integer valued. It is convenient to consider the vicinity of these four peaks/anti-peaks separately. 

Consider first the case when $q$ is a non-negative integer and $\theta$ is close to the peak at $-\frac{\pi}{2}$. Defining the rescaled phase space variables:
\beq 
q = t^2 + t x,\qquad \theta = - \frac{\pi}{2}+ \frac{1}{t}p,
\eeq 
the Wigner function becomes
\beqn 
W(x,p) &=& \frac{1}{ (2\pi)^{3/2}}\int_{-t}^{t+2x}dy\,e^{ip (2y-2x)} e^{-\frac{1}{4}y^2 - \frac{1}{4}(2x-y)^2}\nonumber\\
&=& e^{-\frac{1}{2}x^2-2p^2}\left[\text{erf}(t+x+ip) - \text{erf}(-t-x+ip)\right],
\eeqn 
where
\beq 
\text{erf}(z) = \frac{2}{\sqrt{\pi}} \int_0^z\,dy\,e^{-y^2}.
\eeq
In the $t\to \infty$ limit, the above Wigner function approaches a Gaussian in the rescaled variables, and therefore has no negativity. However, recall that there is a second such peak at $\theta = \frac{\pi}{2}$; for $q \in \mathbb{Z}_{\geq 0}$, this peak comes with an overall positive sign, while for $q\in \mathbb{Z}_{\geq 0}+\frac{1}{2},$ it comes with a negative sign. Thus, in the normalization sum/integral of the Wigner function over the total phase space, the contributions from half-integer $q$ values drop out, while for integer $q$ values, the two peaks at $\theta = \pm \frac{\pi}{2}$ individually contribute $\frac{1}{2}$ each. On the other hand, in computing the negativity, the half-integer peaks do not cancel (because of the absolute value) but add up, and thus we get a total contribution of 2.  Thus, at early times (i.e., sending $\lambda \to 0$ with $t$ held fixed), the time-evolved state in DSSYK exhibits an approximately classical phase where the Wigner negativity stays constant at the value 2.  
\subsubsection*{$\lambda \to \infty$: \text{GUE} limit}
On the other hand, in order to understand the late-time dynamics, we can take the limit $\lambda \to \infty$ while holding $t$ fixed. As we discussed above, in this limit the DSSYK model reduces to the GUE which has constant $b_n$ Lanczos coefficients, with $a_n=0$.  Equivalently, $H_{\text{new}}=\sqrt{\lambda} H_{\text{old}}$ in the $\lambda \to \infty$ limit becomes:
\beq
H=-\left(\alpha+\alpha^{\dagger}\right) \, .
\eeq
As as explained in the previous section, for sufficiently late times the negativity for the GUE  grows as $t^{1/2}$. Thus, at sufficiently late times, we expect Wigner negativity in the DSSYK model to also grow as $t^{1/2}$.  
\subsubsection*{Numerical analysis}
\begin{figure}[t]
    \centering
    \includegraphics[width=0.45\textwidth]{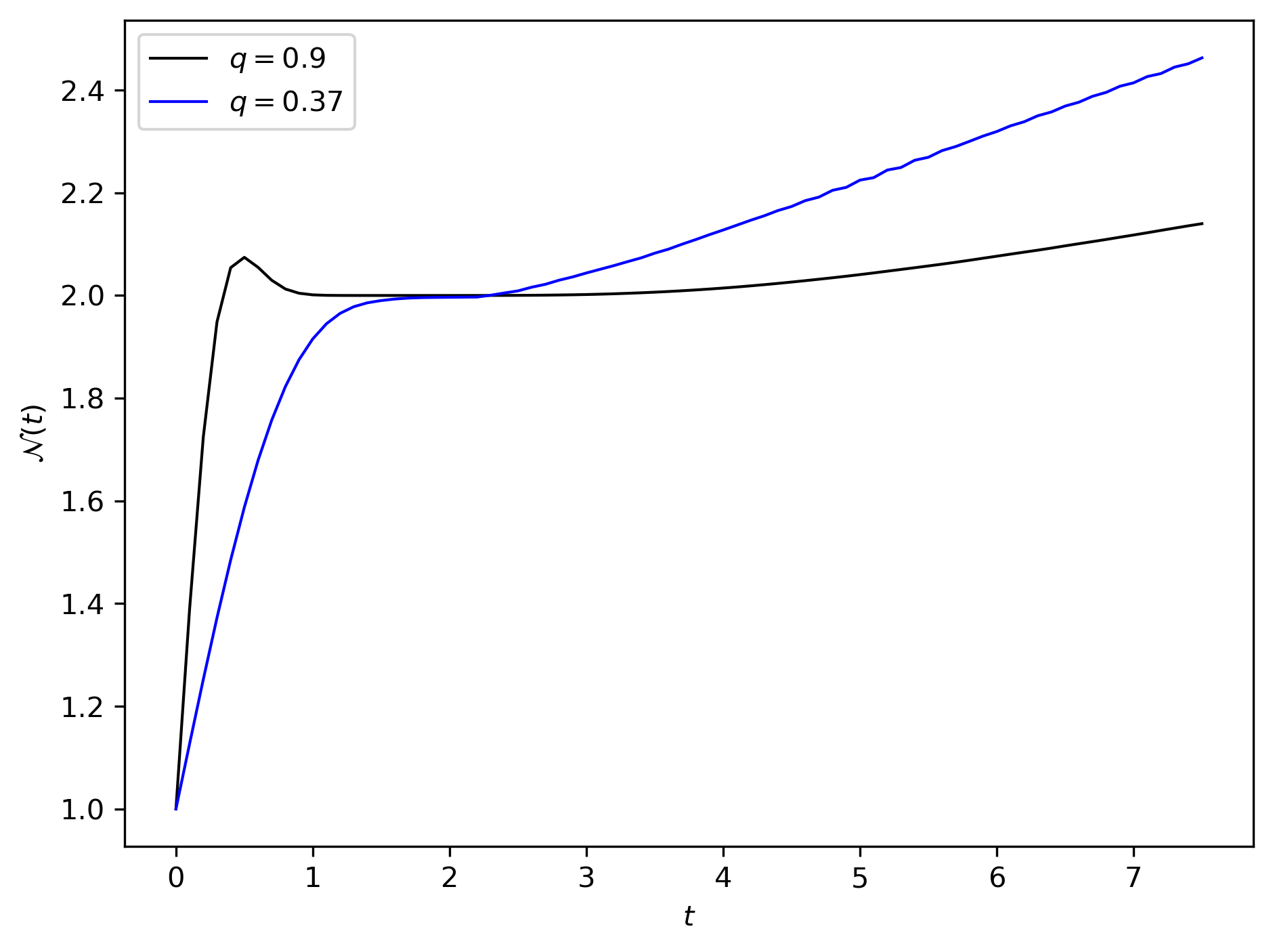}
    \includegraphics[width=0.5\textwidth]{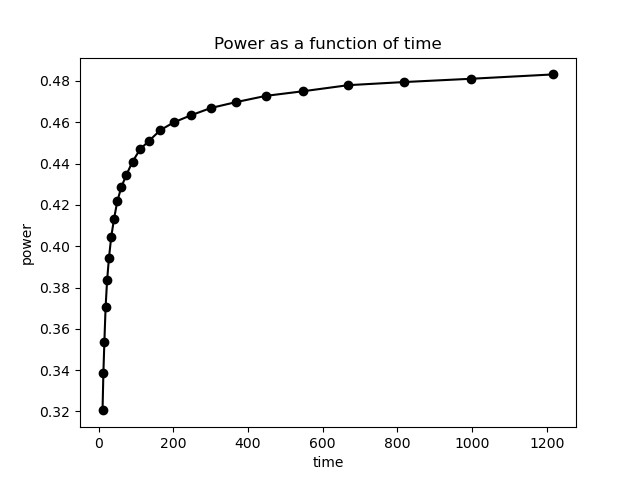}
    \caption{\textbf{(Left)} Wigner negativity in the DSSYK model with no particle insertions for $q=0.37$ (blue) and $q=0.9$ (black); note that the early plateau/saturation phase is longer for a larger value of q. \textbf{(Right)} At late times, the negativity grows as a power law. Here we plot $\gamma = t\frac{d}{dt} \log\mathcal{N}$ as a function of time for $q=0.37$; note that $\gamma$ approximately approaches $1/2$ at late times.}
    \label{fig:SYK0par}
\end{figure}

\begin{figure}[t]
    \centering
    \includegraphics[width=15cm, height=5.5cm]{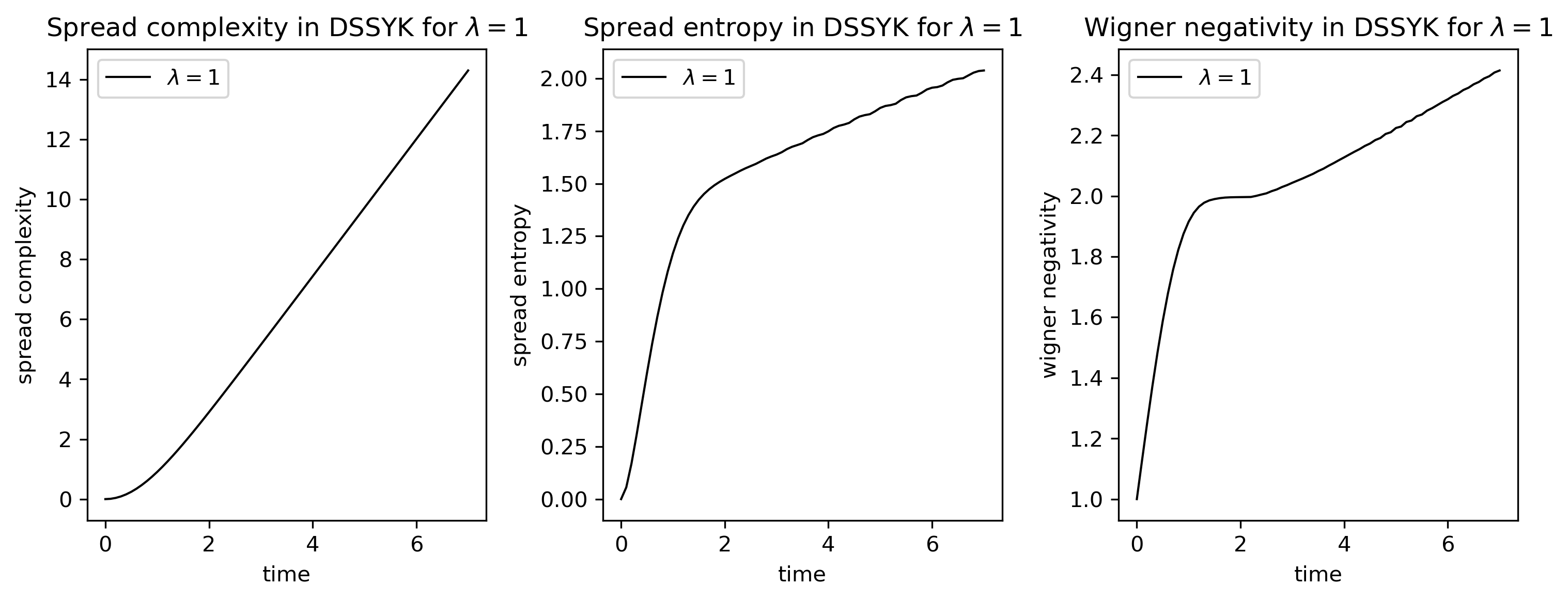}
    \caption{Spread complexity (left), spread entropy (center) and Wigner negativity (right) in the DSSYK model for $\lambda=1$.}  
    \label{fig:DSSYK}
\end{figure}

In Fig.~\ref{fig:SYK0par} we plot Wigner negativity in the DSSYK model for two different values of $\mathfrak{q}$. As expected from the analysis above,  negativity is stable for a while near $\mathcal{N}\sim 2$, and then shows $O(\sqrt{t})$ growth. In Fig.~\ref{fig:DSSYK}, we compare the spread complexity, spread entropy and Wigner negativity for $\lambda=1$. For small $t$, the evolution of the system is effectively that of the Heisenberg model, in this regime  spread complexity grows as $t^2$ \cite{Balasubramanian:2022tpr}, whereas Wigner negativity grows to and then saturates at the value $2$. For large $t$, the system enters the GUE regime. Here the spread complexity grows linearly with time whereas Wigner negativity grows as $\sqrt{t}$.


The discussion in this section has been limited to the $D\to \infty$ limit because we are working in the double scaling limit.  If the Hilbert space dimension $D=e^S$ is finite, we expect that  late time negativity growth should truncate at a time roughly $t \sim e^S$. At this time, the negativity growth should saturate to a value proportional to $e^{S/2}$, indicating the breakdown of the semiclassical, effective description in the Krylov basis. In the case of GUE, this saturation was established numerically in \cite{2024JHEP...05..264B, Basu:2025mmm}; we expect the same behavior in the DSSYK model.   

\section{Epilogue}
In this paper, we have studied the growth of Wigner negativity in Krylov space for various solvable models. In 2d CFTs, we found that the Wigner negativity quickly saturates to an $O(1)$ constant, and that time evolution in the Krylov basis shows emergent classicality (at least for operator dimensions $\Delta > 1/2$). In random matrix theory, we argued on general grounds that the Wigner negativity should grow as $t^{1/2}$ at large but $O(1)$ times, indicating semiclassical time evolution. Finally, in the DSSYK model, we found a classical phase (with constant Wigner negativity) at early times, and a semiclassical phase (with $t^{1/2}$ growth) at late but $O(1)$ times. In all cases, the Krylov basis provides a useful, low-complexity, semiclassical description of quantum dynamics at all $O(1)$ times, and Wigner negativity provides an operationally meaningful probe of this emergent semiclassicality.  We also showed that the growth of spread complexity can be driven by essentially classical dynamics, in that the negativity can remain small even as the spread complexity grows.

We expect that the ideas we have explored in this paper will find natural application in the context of AdS/CFT. The miracle of AdS/CFT is that the bulk gravitational theory provides a semiclassical description of the strongly coupled, chaotic boundary dynamics at large $N$. If we attempt to describe the boundary time evolution of a typical ``black hole'' state (at energy $E\sim O(N^2)$) with respect to a generic choice of basis, then this would look hopelessly complicated on account of the exponential number of states over which the wavefunction would be spread out. Nevertheless, gravity gives a useful semiclassical description of this quantum dynamics, and so it must be that gravity is clever enough to choose the right basis with respect to which the boundary time evolution around certain ``reasonable''\footnote{It would be good to clarify what reasonable here means.} states becomes semiclassical, and remains confined to an $O(1)$ subspace at $O(1)$ times. We have argued in this paper that this is essentially what the Krylov space description also accomplishes, using Wigner negativity as a probe for the emergence of semiclassicality. Our results suggest the compelling picture that gravity should be interpreted as a \emph{low-complexity effective theory}. Indeed, it has long been appreciated \cite{Papadodimas:2013jku, Almheiri:2014lwa} that the gravitational description captures a small ``code subspace'', and the idea that semiclassical gravity can reliably capture low-complexity observables has appeared several times in the literature before, starting from  \cite{Harlow:2013tf} (see \cite{susskind2014computational, Susskind:2014moa, Susskind:2015toa, Brown:2019rox, Kim:2020cds, Balasubramanian:2022fiy, Akers:2022qdl} for an incomplete list of references).  The Krylov space methods used in this work provide a natural, universal language useful in quantifying the dynamics within such code subspaces.

It has become increasingly clear that the subtleties of the large-$N$ limit are intimately tied to the emergence of semiclassical spacetime in AdS/CFT.  In traditional approaches to large-$N$ systems, one  identifies a set of ``collective variables'' in terms of which the path integral can be performed in a saddle-point approximation \cite{Jevicki, Das:1990kaa, Yaffe:1981vf}. However, this method works case-by-case and generally relies on an underlying local description of the path integral. In strongly coupled, chaotic systems -- for instance systems where the Hamiltonian is well-described by a random matrix -- such a local description may not be evident, and so it is a subtle matter to find  collective variables in terms of which time evolution is easy to describe as the system size grows.   Recent work on the existence and properties of large-$N$ limits has focused on properties of von Neumann algebras  \cite{Leutheusser:2021frk, Witten:2021unn}.  Here we are suggesting that the Krylov basis gives a general-purpose recipe for finding  a semiclassical Hilbert space description of dynamics at large $N$.  It would be interesting to understand the connection between these approaches.

\acknowledgments
We would like to thank Chris Akers, Kabir Bajaj, Giuseppe Di Giulio, Abhijit Gadde, Shiraz Minwalla, Mukund Rangamani, Pratik Rath, Joan Simon and Sandip Trivedi for helpful discussions. We are particularly grateful to Kabir Bajaj for early collaboration on this project. OP acknowledges fruitful discussions during the workshop ``Observers, wormholes and complex saddles in cosmology", organized at the Bernoulli Center for Fundamental Studies (EPFL, Lausanne) from 18--22 May 2026. Numerical computations were carried out on the computing clusters at the Department of Theoretical Physics, TIFR, Mumbai. We also thank Ajay Salve and Kapil Ghadiali for their computational support. OP and VS are supported by the Department of Atomic Energy, Government of India, under Project Identification Number RTI-4012 and from the Infosys Endowment for the study of the Quantum Structure of Spacetime. VB was supported in part by the DOE through DE-SC0013528 and the QuantISED grant DE-SC0020360, and thanks ICTS and TIFR for hospitality as this paper was completed. PC is supported by the  Swedish Research Council (VR) under Grant No. 2025-04154 and the ERC Consolidator grant (number: 101125449/acronym: QComplexity). Views and opinions expressed are however those of the authors only and do not necessarily reflect those of the European Union or the European Research Council. Neither the European Union nor the granting authority can be held responsible for them.

\appendix 
\section{Exactly-solvable Lanczos coefficients}\label{App:solvable}
\subsection{The $\text{SL}(2,\mathbb{R})$ model}\label{App:SL2R}
Consider the Lie algebra corresponding to the group $\text{SL}(2,\mathbb{R})$:
\beqn
\commutator{L_1}{L_{-1}}=2L_0, \qquad\commutator{L_0}{L_{\pm 1}}=\mp L_{\pm 1},
\eeqn 
and consider a Hamiltonian of the form:
\beq 
H=\alpha(L_1+L_{-1})+\gamma L_0+\delta 1. 
\eeq
Furthermore, let the initial state $\ket{\psi}$ be the highest weight state of a discrete series representation with scaling dimension $h$. For this choice, the Krylov basis vectors are given by $\ket{K_n}=\ket{h,n}$, where:
\beq
\ket{h,n}=\sqrt{\frac{\Gamma(2h)}{\Gamma(2h+n)n!}} \;L_{-1}^n \ket{h}.
\eeq
Using the $\text{SL}(2,\mathbb{R})$ algebra, one can show \cite{Caputa:2021sib,Balasubramanian:2022tpr} that the Lanczos coefficients in this case are given by:
\beqn
a_n=\gamma(h+n)+\delta,\qquad b_n=\alpha \sqrt{n(2h+n-1)},
\eeqn
and the time-evolved Krylov wavefunction is given by:
\beq
  \psi_n(t)=e^{-i \delta t}\sqrt{\frac{\Gamma(2h+n)}{\Gamma(2 h) \Gamma(n+1)}} g^{2 h} f^n, \label{eq:wvfnc}
\eeq
where:
\beqn
f = \frac{-2 i \alpha}{\mathcal{D}} \frac{\text{tanh}\left(\frac{\mathcal{D} t}{2}\right)}{1+\frac{i \gamma}{\mathcal{D}}\text{tanh}\left(\frac{\mathcal{D} t}{2}\right)},\qquad g = \frac{1}{\cosh(\frac{\mathcal{D} t}{2})+\frac{i \gamma}{\mathcal{D}}\sinh(\frac{\mathcal{D} t}{2})},
\eeqn
with
\beq
\mathcal{D}=\sqrt{4 \alpha^2-\gamma^2}.
\eeq
The return amplitude can be read off from the wavefunction:
\beq
S(t)=e^{i \delta t}\left( \cosh\left(\frac{\mathcal{D} t}{2}\right)-\frac{i \gamma}{\mathcal{D}}\sinh\left(\frac{\mathcal{D} t}{2}\right) \right)^{-2h}. \label{eq:return}
\eeq 

\subsection{The $\text{SU}(2)$ model}\label{App:SU2}
Similar setup governed by the $\text{SU}(2)$ algebra
\be
[J_0,J_\pm]=\pm J_\pm,\qquad [J_+,J_-]=2J_0,
\ee
and evolution of the lowest state $\ket{j,-j}$ with
\be
H=\alpha(J_++J_-)+\gamma J_0+\delta 1,
\ee
has $2j+1$ Krylov basis vectors $\ket{j,-j+n}$, and Lanczos coefficients
\be
a_n=\gamma(-j+n)+\delta,\qquad b_n=\alpha\sqrt{n(2j-n+1)},
\ee
where $n=0,...,2j$.\\
The wave functions can be derived from the BCH formula and read
\be
\psi_n(t)=e^{-i\delta t}\sqrt{\frac{\Gamma(2j+1)}{n!\Gamma(2j-n+1)}}g^{-2j}f^n\,,
\ee
where
\be
f = \frac{-2 i \alpha}{\mathcal{D}} \frac{\text{tan}\left(\frac{\mathcal{D} t}{2}\right)}{1+\frac{i \gamma}{\mathcal{D}}\text{tan}\left(\frac{\mathcal{D} t}{2}\right)},\qquad g=\frac{1}{\cos\left(\frac{\mathcal{D}t}{2}\right)+\frac{i\gamma}{\mathcal{D}}\sin\left(\frac{\mathcal{D}t}{2}\right)}\,,
\ee
with
\be
\mathcal{D}=\sqrt{4\alpha^2+\gamma^2}.
\ee
The return amplitude in this model is
\beq
S(t)=e^{i \delta t}\left( \cos\left(\frac{\mathcal{D} t}{2}\right)-\frac{i \gamma}{\mathcal{D}}\sin\left(\frac{\mathcal{D} t}{2}\right) \right)^{2j}. \label{eq:returnSU}
\eeq 
\subsection{The Heisenberg-Weyl model}\label{App:HW}
For completeness we present the solution governed by the Heisenberg-Weyl algebra
\be
[a,a^\dagger]=1,\quad [N,a^\dagger]=a^\dagger,\quad [N,a]=-a,
\ee
where $N=a^\dagger a$. Consider the evolving Hamiltonian
\be
H=\gamma a^\dagger a+\alpha(a^\dagger+a)+\delta 1.
\ee
If we evolve the vacuum state $\ket{\psi(t)}=e^{-iHt}\ket{0}$, we get infinite set of Lanczos coefficients 
\be
a_n=\gamma n+\delta,\qquad b_n=\alpha\sqrt{n}.\label{LanczosHW}
\ee
The wave functions are
\be
\psi_n(t)=\frac{(-\alpha)^n}{\gamma^n\sqrt{n!}}\left(1-e^{-i\gamma t}\right)^n\exp\left(-i\delta t+\frac{\alpha^2}{\gamma^2}\left(i\gamma t-1+e^{-i\gamma t}\right)\right),
\ee
and the return amplitude in this model is
\be
S(t)=\exp\left(i\delta t+\frac{\alpha^2}{\gamma^2}\left(e^{i\gamma t}-i\gamma t-1\right)\right).
\ee

To gain more intuition, let us consider evolution of the following, more general, states
\be
\ket{\psi(t)}=e^{-iHt}\ket{k}\,,\qquad \ket{k}=\frac{(a^\dagger)^k}{\sqrt{k!}}\ket{0}\,.
\ee
Such states are examples of the so-called displaced number states $D(z)\ket{k}$, obtained by applying a displacement operator $D(z)$ to Fock states, and form a simple and well-controlled class of non-Gaussian states exhibiting pronounced non-classical features. Non-Gaussianity is also a key resource in continuous-variable quantum information, enabling tasks such as universal quantum computation, enhanced metrology, and entanglement distillation beyond the Gaussian regime \cite{walschaers2021non}.

Using the BCH formula
\be
e^{iHt}=e^{\alpha_1(t)}e^{z(t)a^\dagger}e^{-\bar{z}(t)a}e^{i\gamma t\, a^\dagger a}\,,
\ee
with 
\be
z(t)=\frac{\alpha}{\gamma}\left(e^{i\gamma t}-1\right)\,,\qquad \alpha_1(t)=i\delta t+\frac{\alpha^2}{\gamma^2}\left(e^{i\gamma t}-i\gamma t-1\right)\,,
\ee
we can compute the general return amplitude
\bea
S_k(t)&=&\langle k|e^{iHt}|k\rangle=e^{\alpha_1(t)}e^{i\gamma t k}\sum^k_{n=0}\frac{(-\bar{z}(t))^n}{n!}\langle k|e^{z(t)a^\dagger}a^n|k\rangle\nn\\
&=&e^{\alpha_1(t)}e^{i\gamma t k}\sum^\infty_{m=0}\sum^k_{n=0}\frac{z^m(-\bar{z}(t))^n}{m!n!}\sqrt{\frac{k!}{(k-n)!}}\sqrt{\frac{(k-n+m)!}{(k-n)!}}\langle k|k-n+m\rangle\nn\\
&=&e^{\alpha_1(t)}e^{i\gamma t k}k!\sum^k_{n=0}\frac{(-z(t)\bar{z}(t))^n}{(n!)^2(k-n)!}=e^{\alpha_1(t)}e^{i\gamma t k}L_k(|z(t)|^2)\,,
\eea
with $L_k(x)$ being the Laguerre polynomial of degree k. This gives the return amplitude
\be
S_k(t)=e^{i(\gamma k+\delta)t}e^{\frac{\alpha^2}{\gamma^2}\left(e^{i\gamma t}-i\gamma t-1\right)}L_k\left(\frac{4\alpha^2}{\gamma^2}\sin^2\left(\frac{\gamma}{2}t\right)\right)\,.
\ee
This general formula is unfortunately a bit cumbersome for extracting Lanczos coefficients, but we will be mostly interested in cases with $\delta=\gamma=0$. Then, we simply get
\be
S_k(t)=e^{-\frac{1}{2}\alpha^2t^2}L_k(\alpha^2t^2)\,.
\ee
For $k=0$, we have $L_0(x)=1$, and Lanczos coefficients \eqref{LanczosHW} become $a_n=0$ and $b_n=\alpha\sqrt{n}$. The spread complexity grows quadratically \cite{Caputa:2021sib} $C(t)=\alpha^2t^2$.\\
For initial state $\ket{1}$ the return amplitude becomes
\be
S_1(t)=e^{-\frac{1}{2}\alpha^2t^2}(1-\alpha^2t^2)\,,
\ee
and we derive staggered Lanczos coefficients 
\be
a_n=0\,,\qquad b_{2n}=\alpha\sqrt{2n},\qquad b_{2n-1}=\alpha\sqrt{2n+1}\,.
\ee
Consequently, we can derive the solutions of the Schroedinger equation \eqref{eq:SchrEq} for even
\be
\psi_{2n}=\frac{(-1)^n(\alpha t)^{2n}(1+2n-\alpha^2t^2)}{\sqrt{(2n+1)!}}e^{-\frac{1}{2}\alpha^2t^2}\,,\qquad n=0,1,2...\,,
\ee
and odd labels
\be
\psi_{2n-1}(t)=\frac{i(-1)^n(\alpha t)^{2n-1}(1+2n-\alpha^2t^2)}{\sqrt{(2n+1)(2n-1)!}}e^{-\frac{1}{2}\alpha^2t^2}\,,\qquad n=1,2...\,.
\ee
We can verify that
\be
\sum^\infty_{n=0}p_{2n}(t)+\sum^\infty_{n=1}p_{2n-1}(t)=1\,,
\ee
where
\be
p_{2n}(t)=\frac{(1+2n-\alpha^2t^2)^2(\alpha t)^{4n}}{\Gamma(2n+2)}e^{-\alpha^2t^2}\,,\quad p_{2n-1}(t)=\frac{(1+2n-\alpha^2t^2)^2(\alpha t)^{4n-2}}{(2n+1)\Gamma(2n)}e^{-\alpha^2t^2}\,.
\ee
Finally, the spread complexity becomes
\be
C(t)=\sum^\infty_{n=0}2np_{2n}(t)+\sum^\infty_{n=1}(2n-1)p_{2n-1}(t)=\alpha^2t^2+1-e^{-2\alpha^2t^2}\,.
\ee
Clearly, the state $\ket{1}$ has a faster initial growth than $\ket{0}$
\be
C(t)\simeq 3\alpha^2t^2-O(t^4)\,.
\ee
This growth is consistent with $b^2_1t^2$ since now we have $b_1=\sqrt{2}\alpha$.
At late times the evolution becomes quadratic $\sim\alpha^2t^2$, the same as for $\ket{0}$.

\section{Stationary phase condition in random matrix theory} \label{app:stationary}

In this appendix, we solve equation (\ref{eq:stationary_phase}). First notice that since we are restricting to $\theta \in [0,\pi]$, we have $\phi\in[-\frac{\pi}{2},\frac{\pi}{2}]$. To solve it, let $\alpha=\text{cos}^{-1}(y^*)$ and $\beta=\text{cos}^{-1}(2x-y^*)$. Substituting this and taking cosines of both sides we get:
\beqn
\text{cos}(2\phi) &=& \text{cos}(\alpha \pm \beta)\,\nn\\
&=&\text{cos}(\alpha) \text{cos}(\beta) \pm \text{sin}(\alpha) \text{sin}(\beta)\,\nn \\
&=& y^* (2x-y^*) \pm \sqrt{(1-y^{*2}) (1-(2x-y^*)^2))}\,,
\eeqn
where minus comes when $\sigma$ and $\sigma'$ have the same sign and plus comes when they have the opposite signs. This gives a quadratic equation which can be solved to give:
\beq\label{eq:quadratic_sol}
y^*=x \pm \text{tan}(\phi)\sqrt{\text{cos}^2(\phi)-x^2}\,.
\eeq
So now, there are two possible solutions for the stationary phase condition for each $\sigma,\sigma'$, we need to figure out which solution is valid for particular values of $\sigma$ and $\sigma'$.

Moreover, this solution might not be valid for all values of $x, \phi$ (for example, we can see equation (\ref{eq:quadratic_sol}) itself demands $x<\text{cos}(\phi)$ $\forall$ $\sigma,\sigma'$ for $y^*$ to be real). So, we also need to figure out the range of $x$ and $\phi$ for which these solutions are valid. Let's do this case by case:
\paragraph{$\mathbf{\sigma,\sigma'=1:}$}
For this we have, $\text{cos}^{-1}(y^*) -\text{cos}^{-1}(2x-y^*)=2\phi$. Firstly, note that left hand side always decreases when $y^*$ increases, this is true since it's derivative with respect to $y^*$ is always negative, hence only one of the two solutions should be valid. If $\phi>0$, the equation can only be satisfied if $y^*<x$ and if $\phi<0$ then $y^*>x$, this implies that the only possible solution is $y^*=x - \text{tan}(\phi)\sqrt{\text{cos}^2(\phi)-x^2}$. Finally, since $\text{cos}^{-1}(y^*) -\text{cos}^{-1}(2x-y^*)$ is a decreasing function $2 \phi$ must lie between it's end point values. Now, if $x<\frac{1}{2}$ then $0<y<2x$, so the endpoint values are, $\phi=\frac{\pi}{4}-\frac{\text{cos}^{-1}(2x)}{2}$ and $\phi=-\frac{\pi}{4}+\frac{\text{cos}^{-1}(2x)}{2}$. Similarly, if $x>\frac{1}{2}$ then $2x-1<y<1$, so the endpoint values are $\frac{\text{cos}^{-1}(2x-1)}{2}$ and $-\frac{\text{cos}^{-1}(2x-1)}{2}$. In conclusion, for this case the stationary point exists for $x<1/2$ when $-\frac{\pi}{4}+\frac{\text{cos}^{-1}(2x)}{2}<\phi<\frac{\pi}{4}-\frac{\text{cos}^{-1}(2x)}{2}$ and for $x>1/2$ when $-\frac{\text{cos}^{-1}(2x-1)}{2}<\phi<\frac{\text{cos}^{-1}(2x-1)}{2}$. 
\paragraph{$\mathbf{\sigma,\sigma'=-1:}$}
For this case the stationary point condition is, $-\text{cos}^{-1}(y^*) +\text{cos}^{-1}(2x-y^*)-2\phi=0$. This case is very similar to our previous case, in fact it is symmetric under $y \to 2x-y$. Using this we can say that the solution is $y^*=x + \text{tan}(\phi)\sqrt{\text{cos}^2(\phi)-x^2}$ and it exists for $x<1/2$ when $-\frac{\pi}{4}+\frac{\text{cos}^{-1}(2x)}{2}<\phi<\frac{\pi}{4}-\frac{\text{cos}^{-1}(2x)}{2}$ and for $x>1/2$ when $-\frac{\text{cos}^{-1}(2x-1)}{2}<\phi<\frac{\text{cos}^{-1}(2x-1)}{2}$.
\paragraph{$\mathbf{\sigma=1, \sigma'=-1:}$}
The stationary point condition in this case is given by, $\text{cos}^{-1}(y^*) +\text{cos}^{-1}(2x-y^*)=2\phi$ This is different from the previous two cases as the left hand side is neither decreasing nor increasing throughout the range and instead has a turning point, this can be seen by taking it's derivative w.r.t. y, the derivative is:
\beq
-\frac{1}{\sqrt{1-y^2}}+\frac{1}{\sqrt{1-(2x-y)^2}}\,,
\eeq
and is negative for $y>x$ and positive for $y<x$ with $y=x$ being the turning point. Now, suppose $x<1/2$ so that $0<y<2x$, for $y<x$ the possible solution is $y^*=x - \text{tan}(\phi)\sqrt{\text{cos}^2(\phi)-x^2}$ with the range of validity being $\frac{\pi}{4}+\frac{\text{cos}^{-1}(2x)}{2}<\phi<\text{cos}^{-1}(x)$. Similarly, for $y>x$ it is $y^*=x + \text{tan}(\phi)\sqrt{\text{cos}^2(\phi)-x^2}$ with the range of validity being the same. \\
Whereas, if $x>1/2$, so that $2x-1<y<1$, for $y<x$ the possible solution is $y^*=x - \text{tan}(\phi)\sqrt{\text{cos}^2(\phi)-x^2}$ with the range of validity being $\frac{\text{cos}^{-1}(2x-1)}{2}<\phi<\text{cos}^{-1}(x)$. Similarly, for $y>x$ it is $y^*=x + \text{tan}(\phi)\sqrt{\text{cos}^2(\phi)-x^2}$ with the range of validity being the same.
\paragraph{$\mathbf{\sigma=-1,\sigma'=1:}$}
This case is analogous to the previous case after replacing $\phi \to -\phi$. \\

\bibliographystyle{JHEP}
\bibliography{Reference_wigner.bib}
\end{document}